\documentclass[manuscript]{acmart}
\usepackage{xspace}
\usepackage[normalem]{ulem}

\newenvironment{interfacefont}{\fontfamily{lmtt}\selectfont}{\par}

\newcommand{\sysname}{AI Threads\xspace}

\newcommand{\mchat}{{\interfacefont{Main Chat}}\xspace}
\newcommand{\thread}{{\interfacefont{Thread}}\xspace}
\newcommand{\ddict}{{\interfacefont{Data Dictionary}}\xspace}

\newcommand{\arrowleft}{%
  \begingroup\normalfont
  \includegraphics[height=\fontcharht\font`\B]{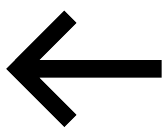}%
  \endgroup
}

\newcommand{\arrowright}{%
  \begingroup\normalfont
  \includegraphics[height=\fontcharht\font`\B]{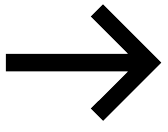}%
  \endgroup
}

\newcommand{\lucky}{%
  \begingroup\normalfont
  \includegraphics[height=1.25\fontcharht\font`\B]{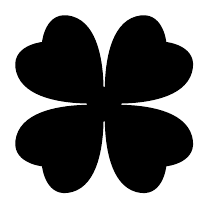}%
  \endgroup
}

\newcommand{\ithread}{%
  \begingroup\normalfont
  \includegraphics[height=1.75\fontcharht\font`\B]{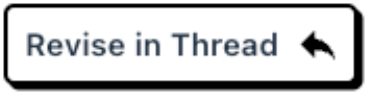}%
  \endgroup
}

\newcommand{\redo}{%
  \begingroup\normalfont
  \includegraphics[height=1.75\fontcharht\font`\B]{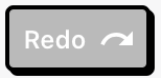}%
  \endgroup
}

\usepackage{listings}
\usepackage{wrapfig}
\usepackage{multirow} 
\usepackage{array}
\usepackage{booktabs}
\usepackage{fancybox}
\usepackage{algorithm}
\usepackage{fontawesome}

\usepackage{listings}
\usepackage{xcolor}
\usepackage{framed}

\definecolor{color1}{RGB}{117, 112, 179}
\definecolor{color2}{RGB}{27, 158, 119}
\definecolor{color3}{RGB}{217, 95, 2}
\definecolor{color4}{RGB}{102, 102, 102}
\definecolor{color5}{RGB}{230, 171, 2}

\definecolor{question}{RGB}{52, 88, 235}
\definecolor{response}{RGB}{235, 143, 25}

\begin{document}

\title{Conversational AI Threads for Visualizing Multidimensional Datasets}


\author{Matt-Heun Hong}
\affiliation{%
  \institution{University of North Carolina}
  \country{USA}}
\email{mhh@cs.unc.edu}

\author{Anamaria Crisan}
\affiliation{%
  \institution{Tableau Research}
  \country{USA}}
\email{acrisan@tableau.com}

\renewcommand{\shortauthors}{Hong et al.}

\begin{abstract}

Generative Large Language Models (LLMs) show potential in data analysis, yet their full capabilities remain uncharted. Our work explores the capabilities of LLMs for creating and refining visualizations via conversational interfaces. We used an LLM to conduct a re-analysis of a prior Wizard-of-Oz study examining the use of chatbots for conducting visual analysis. We surfaced the strengths and weaknesses of LLM-driven analytic chatbots, finding that they fell short in supporting progressive visualization refinements. From these findings, we developed \sysname, a multi-threaded analytic chatbot that enables analysts to proactively manage conversational context and improve the efficacy of its outputs.  We evaluate its usability through a crowdsourced study ($n=40$) and in-depth interviews with expert analysts ($n=10$). We further demonstrate the capabilities of~\sysname on a dataset outside the LLM's training corpus. Our findings show the potential of LLMs while also surfacing challenges and fruitful avenues for future research.

\noindent\textbf{Code \& Supplemental Materials:} \url{https://osf.io/6wxpa/}

    
\end{abstract}




\begin{teaserfigure}
    \includegraphics[width=\textwidth]{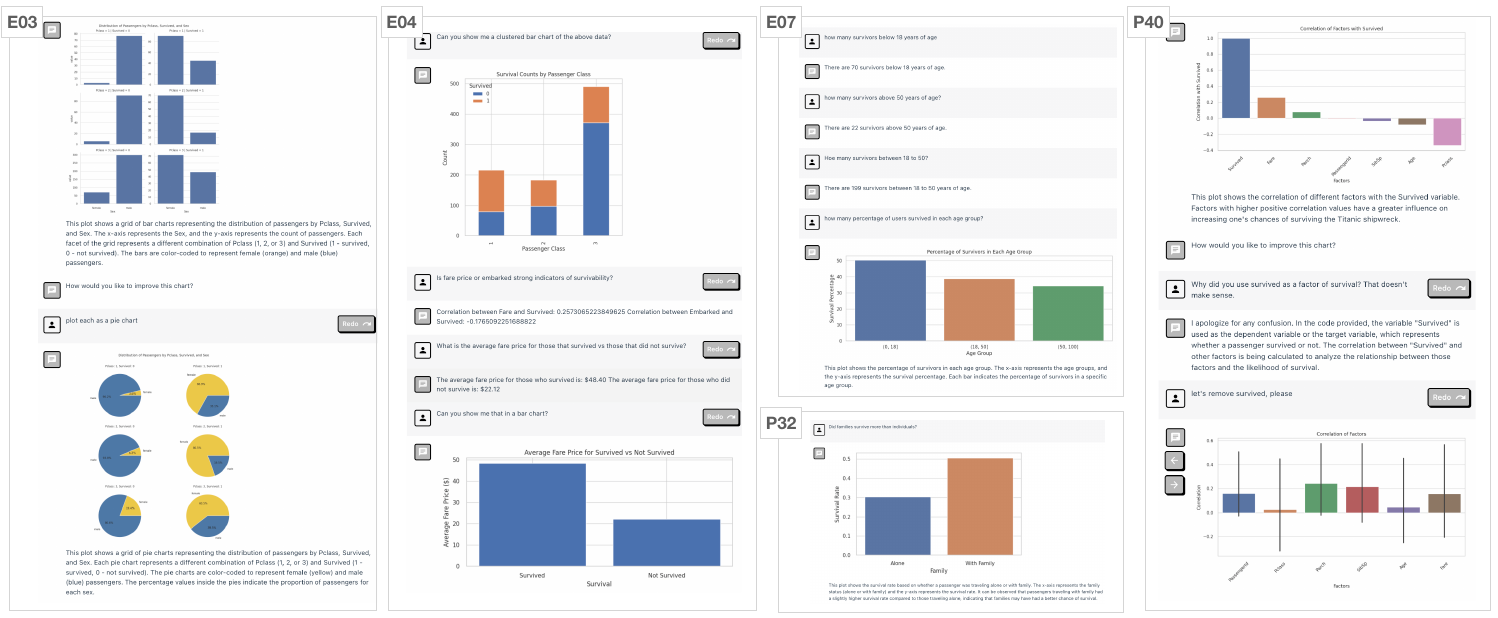}
    \caption{\textbf{Excerpts of participants' analytic conversations with~\sysname.} We show participant conversations in threads modifying the visual encoding (E03, E04, P40), as well as for sequential (E07) and individual (P32) utterances.  Excerpts were derived from participants in crowd-sourced (P\#) and interview (E\#) studies that we conducted. \sysname is capable of editing operations that add or remove attributes to encodings and change encoding types, among others. Overall, the chatbot proved itself capable, but we still encountered instances of incorrect responses, including within a continuous thread (P40).}
    \Description{}
    \label{fig:participant_teaser}
\end{teaserfigure}

\maketitle

\section{Introduction}

Generative Pre-trained Transformer (GPT) large language models (LLMs)~\cite{radford:gpt2:2019} can assist in data-driven knowledge work by generating analytic code from natural language utterances~\cite{Feng:CodeBert:2020,Austin:ProgramSynth:2021,Poesia:Synchromesh:2022}. These capabilities can support novice and expert analysts alike by reducing barriers to conducting preliminary investigations of data~\cite{Maddigan:Chat2Vis:2023,Dibia:LIDA:3034,cheng:llm_analyst:2023}.
~However, LLMs do not reflect the first attempts to facilitate analyses through natural language. Prior research on Visual Natural Language Interfaces (V-NLIs) and conversational agents have explored varying techniques for translating plain language utterances into corresponding data transformations and visualization specifications~\cite{Hoque:ChartQAStar:2022,Shen:NLISurvey:2023,mctear:conversational_agent:2022,voigt:VNLISurvey:2022,Srinivasan:NLI:2017}. A recent head-to-head comparison of LLMs and several V-NLIs suggests that LLM performance is as capable, and potentially superior, to prior systems for resolving ambiguities around user intent behind an utterance~\cite{Maddigan:Chat2Vis:2023}. However, this investigation was shallow and did not probe the potential for LLMs to operate as part of conversational interfaces for more dynamic analytic dialogues. As such, the capabilities of LLMs for conducting data analyses remain uncharted.

Analytic conversations have a complex structure that includes shifting conversation context~\cite{tory:tc17:2019} and sequential utterances that progressively modify analyses and visual encodings~\cite{Fast:Iris:2019,Srinivasan:NLI:2017}. Here, we explore how LLMs handle this shifting conversational context.  
Our research has three parts. First, we re-examine the results of a prior Wizard-of-Oz~\cite{tory:tc17:2019} study and summarize key findings on how an LLM-based analytic agent responds. A key feature of ~\cite{tory:tc17:2019} is that they explicitly mark the shifts in conversational context, which we use as a ground truth in our assessments. Next, we developed~\sysname, a conversational interface that uses an LLM as an AI agent to produce both visual and textual outputs in response to user utterances. A novel aspect of~\sysname is the incorporation of multiple conversational threads, a common feature of messaging apps, to manage the context provided to the AI agent. In the third part, we use the data and objectives from~\cite{tory:tc17:2019} to conduct two usability studies that assessed the validity, robustness, and utility of~\sysname. The first is a crowd-sourced study with 40 participants, followed by a second in-depth study that conducted semi-structured interviews with 10 analytic experts. The findings from these user studies show that participants had an overall positive perception of~\sysname and that they found it to be quite capable. However, full reviews of the session logs also revealed its pitfalls and limitations.  Expert participants provided further insight into the use of AI chatbots relative to their existing analytic processes.

Collectively, our research presents the following three contributions. \textbf{First, we discuss the structure of multi-threaded visual analytic conversations and describe their significance for LLM AI Agents.} We demonstrate how this contribution can help to proactively manage conversational context and improve the organization of the analysis as it progresses.  \textbf{Second, we present the design and implementation of \sysname}. We ground its development of in four design goals that aim to enable natural conversational flows for data analysis. \textbf{Finally, we present a set of user studies that showcase both the capabilities and limitations of LLM as analytic conversational agents.} We also identify how such conversational agents stand to influence data analysis.

As conversational agents supported by LLMs are poised to play a larger role in data analysis, including the generation of visualizations, our research presents a novel approach to supporting these complex and multi-faceted conversations. 


\section{Related Work}\label{rw}

We summarize related work pertinent to Visual Natural Language Interfaces, Conversational agents, and visualization recommendation systems. 

\subsection{Visual Natural Language Interfaces (V-NLIs)}\label{rw:nli}
 
V-NLIs have undergone an evolution of techniques to understand utterances to create appropriate mappings to dataset attributes, data transformation, and visual encodings. ~\cite{Maddigan:Chat2Vis:2023,Shen:NLISurvey:2023}

Initial approaches, such as DataTone~\cite{Gao:DataTone:2015}, Eviza~\cite{setlur:eviza:2016}, and FlowSense~\cite{Yu:FlowSense:2020} used parse trees, grammar, rules, or heuristics to resolve these mappings. These prior works also uncovered the challenges of managing ambiguous utterances and capturing and modeling the semantics, intent, and pragmatics of these conversations~\cite{Hoque:Evizeon:2018}. The Orko system~\cite{Srinivasan:orko:2018} reified these channels with a non-tabular dataset as well. While these techniques are still used in contemporary V-NLI systems~\cite{Shen:NLISurvey:2023}, they are also critiques as being too brittle, which limits their efficacy~\cite{Maddigan:Chat2Vis:2023}.

More recent approaches have tried to learn these mappings directly from the data using machine learning approaches, including Deep Neural Networks. Examples for such systems include, Advisor~\cite{Liu:ADVISor:2021}, ncNet~\cite{Yuyu:ncNet:2022}, Talk2Data~\cite{Guo:Talk2Data:2021},  and Urania~\cite{Guo:Urania:2023}. DNN approaches vary with regards to the specific architecture and labeled training corpora they use to learn mapping functions. 
More recently, the use of LLMs has been explored for V-NLIs by leveraging their program generating capabilities ~\cite{Austin:ProgramSynth:2021} to generate and execute code for transforming and visualizing data. 
Unlike DNN approaches, LLMs 
may only require prompt engineering and a few labeled examples to generate visualizations from natural language queries.
As in the current work, Chat2Vis~\cite{Maddigan:Chat2Vis:2023} uses GPT-3.5 to generate visualizations from natural language queries of a dataset. 
Their preliminary evidence suggests LLMs outperform other approaches for understanding a natural language utterance and producing an appropriate visualization.
This transition from many early attempts at V-NLIs to the use of LLMs follows an arc from heuristic (e.g.,~\cite{setlur:eviza:2016,Gao:DataTone:2015}) to data-driven approaches that many fields adjacent to NLP have similarly experienced.

We extend the prior research by conducting a deeper investigation into the use of LLMs to support the visual analysis of data.  Compared to Maddigan~\textit{et al.} we conduct a more comprehensive assessment of a LLM's capabilities with data from actual human subjects, not just scenarios. Moreover, the~\sysname system we develop addresses many of the limitations that they identify. 

\vspace{-3mm}
\subsection{Conversational Agents for Visual Data Analysis}\label{rw:convo_agent}

Conversational agents, or chatbots, facilitate an interactive dialogue between the analyst and an automated agent~\cite{mctear:conversational_agent:2022}.  V-NLI research has explored the potential of conversational interfaces powered by chatbots to conduct a data analysis~\cite{Kavax:ChatvNLI:2023,Setlur:gricean_maxims:2022}. Iris~\cite{Fast:Iris:2019}  uses a single-threaded conversational interface that combination of multiple utterances to produce an appropriate response; notably, Iris produces both textual and visual responses.  
Kim~\textit{et. al.}~\cite{Kim:ChatQA:2020} implement a pipeline that leverages natural language templates to facilitate a conversation for explaining visualizations.  Both~\cite{Fast:Iris:2019} and~\cite{Kim:ChatQA:2020} touch upon the importance of composition operations (i.e. requiring multiple operations) to carry out these analytic conversations.  Finally, to support participants in visual analysis, SNOWY~\cite{Srinivasan:Snowy:2021} produces ranked suggestions for utterances that users may pose of the data and visualization. 

Adjacent to these efforts of system development, prior research has also explored end-users expectations for analytic chatbots and the structure of analytic conversations. Hearst and Tory~\cite{Hearst:ChartChat:2019} show when conversing with analytic chatbots participants sometimes show a greater preference for text responses over visualizations. Setlur and Tory~\cite{tory:tc17:2019} use a Wizard-of-Oz  study to study the relationship between participant utterances and the automatic generation of visualizations, including the refinement of visualizations. They summarize their findings as a state transition model and discuss the importance of context and intent to facilitate these transitions and support both conversational and visual coherence. In follow-up research, they also demonstrate the importance of the input medium and modalities (voice or text) in influencing these conversations~\cite{Setlur:gricean_maxims:2022}. They found that text interactions in combination with conversational threading supported a deeper analysis compared to other modalities.

The utility of LLMs as analytic chatbots has not been extensively explored. Moreover, recent research has shown that LLMs can struggle with compositionality~\cite{Dziri:Compositionality:2023}. We present a detailed examination of LLMs for conducting analytic conversations that include compositional tasks.

\vspace{-2mm}
\subsection{Visualization Recommendation}\label{rw:vizrec}
Visualization recommendation (VisRec) techniques attempt to automatically generate visualizations primarily based on the attributes of the data. Unlike V-NLIs they are not tied specifically to natural language inputs, but have overlapping techniques to generate visualizations from user inputs. 

Several techniques use both rule-based and heuristic approaches to general visualizations. ShowMe~\cite{Mackinlay:ShowMe:2007} uses a rule-based approach predicated on the field types (categorical or numeric) and best practices ranking. It requires the use of specifying one or more fields before it can automatically generate an encoding. VizDeck~\cite{Key:VizDeck:2012} similarly used data properties to recommend interactive dashboards, instead of single encodings. ShowMe was further extended in the Voyager and Voyager2 systems~\cite{Wongsuphasawat:Voyager:2016}, which use statistical properties of the dataset to automatically generate and rank visualizations of a dataset while relaxing the requirement of an initial user specification. Draco~\cite{Moritiz:Draco:2019} presented an approach to automatically incorporate perceptual efficacy experiments into these visualization recommendation paradigms. ChartSeer~\cite{Zhao:ChartSeer:2022}, Chart2Vec~\cite{Chen:Chart2Vec:2023}, and DeepEye~\cite{Luo:DeepEye:2018} propose machine learning approaches as alternatives to the rule-based methods.  Both ChartSeer~\cite{Zhao:ChartSeer:2022} and Chart2Vec~\cite{Chen:Chart2Vec:2023} learning a vector representation of visual encodings, albeit in different ways, to recommend visualizations, while DeepEye~\cite{Luo:DeepEye:2018} uses a learning-to-rank approach. Recently, LIDA was developed as a way to recommend visualizations by using an LLM to generate specifications~\cite{Dibia:LIDA:3034}. 

The previously mentioned approaches focus largely on specifications of the data attributes and mapping these to visual encodings.  Other techniques can take additional contextual factors such as the analytic task, user intent, prior visualizations, and even the domain application. TaskVis~\cite{Agus:TaskVis:2021} uses a pre-defined set of tasks and a rule-based approach to recommend visualization from a combination of data fields and analytic tasks. Frontier~\cite{Lee:Frontier:2022} similarly uses pre-defined task and intent classifications for data analysis to generate and organize visualization recommendations. The concepts of Frontier are reified in the Lux~\cite{Lee:Lux:2021}, a Jupyter notebook widget that interacts with Pandas, a widely used Python data analysis library. Medely extends these intent-based recommendations further to generate dashboard visualizations~\cite{Pandey:Medely:2023}. GraphScape~\cite{Kim:GraphScape:2017} can propose modifications to existing encodings using a graph transition model. Finally, visualization recommendations can also take domain context into account. For example, GenoRec~\cite{Pandey:GenoRec:2023} and GEViTRec~\cite{Crisan:GEviTREc:2022} both leverage data from the genomics domain to generate recommendations.

Our research combines many of the ideas from several disparate visual recommendation efforts, notably, using LLM (as LIDA does), visual sequencing to modify encodings, and the incorporation of contextual factors.

\section{Re-evaluating Analytic Conversations with AI Chatbots}\label{formative_study}

We conducted a re-analysis of study by Setlur and Tory~\cite{tory:tc17:2019} that examined the dialogue of  visual analytic conversations. Our goal is not to reproduce the precise visualizations from their Wizard-of-Oz (WoZ) study, but instead, to evaluate~\textit{if} and~\textit{how} a LLM is able to handle transitions in conversations from modify an existing visualization or generate a new one.  We summarize the key findings here and provide a deeper analysis in Appendix~\ref{appendix_formative_study} at the end of this manuscript.


\subsection{Overview of Tory and Setlur (2019)}

Tory and Setlur~\cite{tory:tc17:2019} conducted a WoZ study to examine analytic conversations with an AI chatbot. Participants were tasked to explore the canonical Titanic dataset and answer the question \textit{``What characteristics made it more likely that a passenger survived?''} Participants conducted their analysis by entering utterances (textual inputs) via a research prototype of Tableau's Ask Data feature. The Wizard generated a visualization in response and these were displayed to the participant. The study had four conditions. Two conditions treat each utterance as independent and produce a unique visualization for each, and two that take into account the previous visualization -- these were called the context conditions. In the two context conditions, participants could elect to modify an existing visualization or begin a new analysis. The retention or shift in context was either explicitly stated (e.g., uttering ``retain'', or ``start over'') or implicitly determined (e.g., uttering ``add to this'', ``remove this''). The Wizard would respond to such utterances according to the prescribed rules of the study. 

We sought to \textit{``replay''} the conversations from the context conditions using an LLM. The retention or shift in context observed in Tory and Setlur~\cite{tory:tc17:2019}  delineate discourse chunks that split the analytic conversation between utterances signifying new analysis questions or goals and those intended to modify an existing visualization (see Appendix~\ref{app:disourse_chunks} and Appendix~\ref{app:analytic_convo_evo}). We use these contextual transitions as ground truth to evaluate an LLM. To further examine the LLM's performance we also annotated utterances from this according to three phrasal classifications  (interrogative, imperative, declarative)~\cite{Craige:discourse:2018,kuijpers:mmi2:1992} and assessed the ambiguity of these utterances according to the criteria proposed by Srinivasan~\textit{et al.}~\cite{Srinivasan:Utterance:2021} (see Figure~\ref{app:utterance_classification} and Appendix~\ref{app:utterance_classification}). 

\subsection{Replication with a ADA}\label{formative:chatgpt}

To conduct this evaluation, we used OpenAI's ChatGPT Advanced Data Analysis (ADA) plugin\footnote{We used the version of ADA released before July 15, 2023}. 
ADA is an extension of ChatGPT for code generation in Python and is capable of conducting analytic conversations\footnote{ADA was initially released as the ChatGPT Code Interpreter Plugin~\url{https://openai.com/blog/chatgpt-plugins\#code-interpreter}}. Managing the conversation context for an LLM is a nascent area of development. Currently, a tool like ADA takes an entire conversation as context -- up to a set token limit. Once the token limit is reached, there are different strategies for managing context including summarizing the previous conversation, performing an embedding search to identify relevant parts of the conversation, or actively filtering prior utterances. We use ADA's default context management. Some or all of the aforementioned strategies and likely others may already be incorporated into ADA, but this information is not publicly available at present. 

\begin{figure}[t]
    \centering
    \includegraphics[width=1\textwidth]{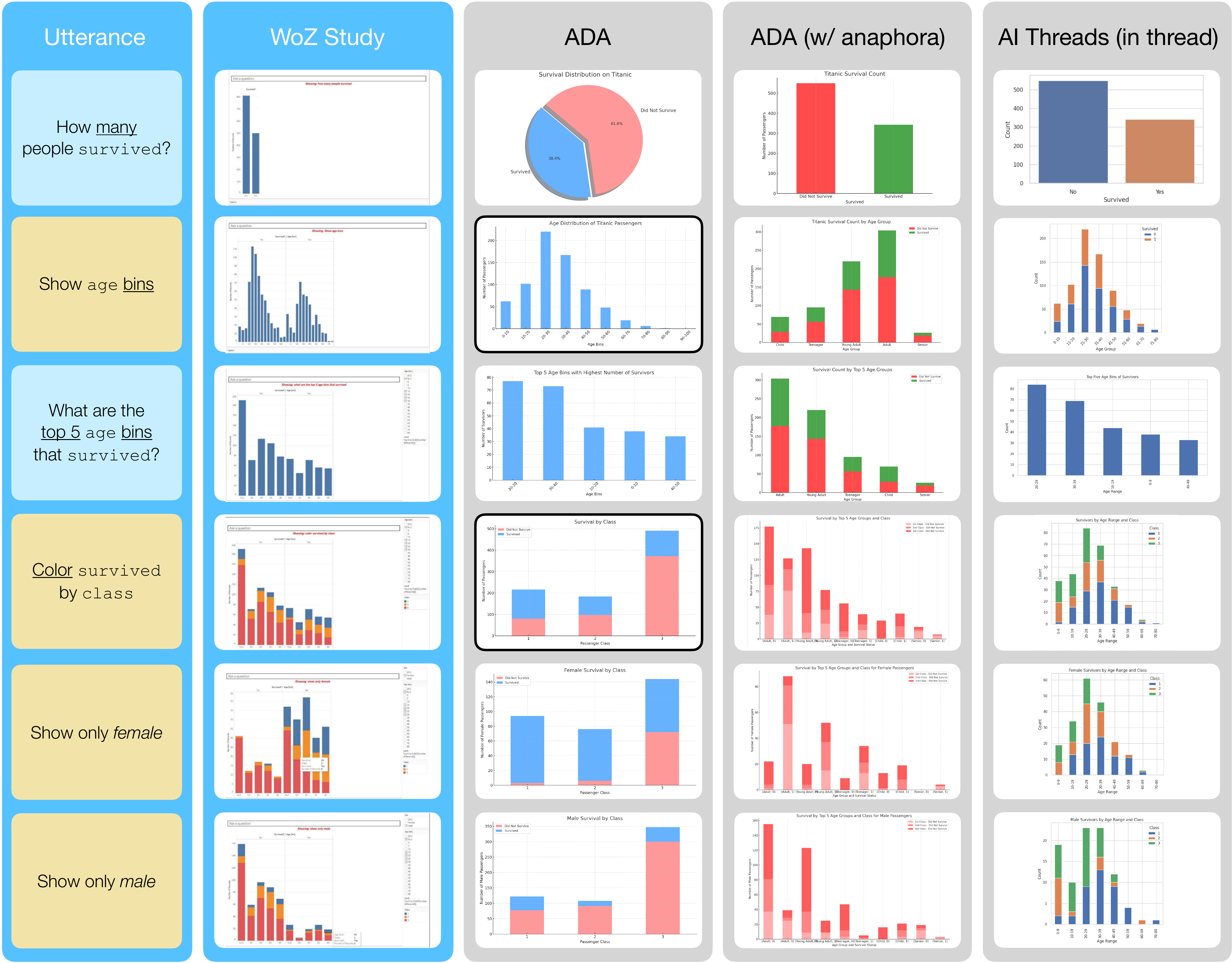}
    \caption{\textbf{Comparison for visualization produced between the WoZ study and different systems.} Progressive utterances for modifying an initial visual encoding (top row) by different systems. Black boxes around the visualization indicate the creation of a new visualization---a failure to apply a progressive update when given an utterance with the intent to modify a previous visualization.}
    \label{fig:comparison}
\end{figure}
\subsubsection{Findings}\label{formative:chatgpt_findings}
A summary of the findings  between the original WoZ study and ADA is summarized in Figure~\ref{fig:comparison}. The default output for ADA is a textual, not visual, response but this behavior can be modified with the appropriate prompts (see Appendix~\ref{app:chatgpt_repeat}). Overall, we found that ADA was able to correctly reproduce the initial data visualization regardless of the attribute, data, or visual ambiguity of the utterance. For example, the interrogative utterance~\textit{``What percentage survived by age in 5-year bins?''} specifies the target attributes explicitly (survived and age), the data transformations (five-year bins and aggregation to a percentage), and leaves it to ADA to choose the visual encoding. ADA produces the appropriate bar chart. We observed that it can also respond effectively to non-explicit attribute references (e.g., \textit{``show only male''} or \textit{``how much did each passenger pay to board the Titanic''}).  Where we observed ADA struggle was transition context without an explicit reference to the previous visualization.

We observed that many of the modifying utterances in~\cite{tory:tc17:2019} were terse and either imperative or declarative (e.g.,~\textit{``class color by fare bin''}, \textit{``add trend line''}, or \textit{``remove class''}; see also Appendix~\ref{app:analytic_convo_evo}). When an utterance had explicit anaphoric references, such as \textit{this} or \textit{add}, ADA was typically able to resolve the contextual dependency and modify the previous visualization. However, when there was no anaphoric reference, such as in the example \textit{``class color by fare bin''}, ADA interpreted this utterance as independent and generated a new visualization even though the participant intended to modify the previous visualization.  However, ADA's behavior was also inconsistent. After several anaphoric references, ADA appeared to learn that the participant intended to modify the previous visualization and interpreted subsequent utterances as intending to modify the previous visualization. This interpretation was not always correct and in practice could frustrate an end-user, as Tory and Setlur also observed~\cite{tory:tc17:2019}. Using utterances to start a new visualization (e.g., \textit{``start over''}) also had unpredictable side effects because it was not clear to ADA at which point it should restart the analysis. For example, its default response would be to reload the entire dataset instead of just beginning a new visualization. It was necessary to be explicit to produce an appropriate output (e.g., ~\textit{``I am done modifying the previous visualization, let's start a new one''}). 

Overall, we often needed to substantively modify the utterances from~\cite{tory:tc17:2019} to make them less ambiguous by adding anaphoric references to produce an appropriate response. This was a trial-and-error process and did not always produce the desired results. Additionally, the conversational thread became difficult to follow because it contained failed attempts to modify the utterance. Anaphoric references, while useful, are a manual and~\textit{ad hoc} approach for an end-user to manage context for the chatbot. While they are workable, they significantly impact the overall coherence of the conversation. 

\subsection{Implications: AI agents and analytic conversations}
\textbf{Analysts do not precisely state their analytic goals at the outset, iterative refinement is the norm.} Our re-evaluation of Tory and Setlur~\cite{tory:tc17:2019} identifies several strengths and limitations of LLMs serving as conversational analytic agents. While we observed that ADA appeared to perform well at composing a singular visualization, it struggled to effectively manage a transitioning conversational context that is common in analytic dialogue. Conversational agents are developed for many purposes~\cite{mctear:conversational_agent:2022}, the structure and nuances of analytic conversations merit their own set of unique approaches.


\section{Conversational AI Agent Design Goals}\label{design_goals}

We draw on the findings from our re-evaluation of Tory and Setlur~\cite{tory:tc17:2019}, the structure of analytic conversations (Section~\ref{rw:convo_agent}), V-NLI and conversational interfaces (Section~\ref{rw:nli}), and finally research on analytic notebooks~\cite{Kery:NotebookStory:2018,Adam:Explore_Notebooks:2018}
to derive a set of design criteria for an analytic conversational agent.

\vspace{1mm}
\noindent\textbf{DG1: Support Conversational Coherence.}\label{coherence} The iterative nature of data analysis can result in a complex dialogue. Commonly this analytic dialogue is found not in conversational interfaces, but typically organized and presented within computational notebooks~\cite{Liu:Notebooks_Alt:2020,Kery:NotebookStory:2018,Adam:Explore_Notebooks:2018}. These notebooks interplay visualization, code, and text annotations that analysts structure over time into a more coherent analysis. In aggregate, this refinement process results in a branching structure~\cite{Deepthi:NotebookCodeEvo:2023,Kery:NotebookStory:2018,Adam:Explore_Notebooks:2018} that resembles the conversational evolution and contextual shifts we observed in~\cite{tory:tc17:2019} (see Appendix~\ref{app:analytic_convo_structure} and~\ref{app:analytic_convo_evo}). Transitioning a linear conversational flow, such as is used in ADA, to a multi-threaded one that resembles this branching structure will improve conversational coherence and consequently improve user experience. Prior research with chatbots~\cite{Setlur:gricean_maxims:2022} also shows that participants prefer this non-linear structure as well. 

\vspace{1mm}
\noindent\textbf{DG2: Enable Naturalistic Conversation.}\label{naturalistic} 
Prior research for V-NLIs and conversational agents specifies several criteria for effectively supporting naturalistic conversations. These include being able to resolve utterance ambiguities and incompleteness, sequential utterances, in addition to understanding the user's analytic intent (see Section~\ref{rw:nli}). In response to end-user utterances, the chatbot should seamlessly support both textual and visual as appropriate~\cite{Hearst:ChartChat:2019,Fast:Iris:2019}. Finally, should the circumstance arise, the chatbot should fail gracefully with feedback to the user~\cite{Amershi:HAI:2019}. Elements of this design goal have already been broadly articulated in the literature and represent essentially table stakes for an analytic conversational agent. While prior systems have made progress on many of these goals there remain open challenges that are introduced by the combination of the above factors, the need to support conversational coherence, and the properties of LLMs.
%

\vspace{1mm}
\noindent\textbf{DG3: Allow Analysts to Proactively Manage Context.}\label{context-mgmt} 
Prior conversational V-NLIs use varying techniques for managing the state and conversational shift~\cite{mctear:conversational_agent:2022}, but the novelty of using an LLM for data analysis creates new context management challenges.
Our findings suggest that managing context for analytic conversations may require new approaches compared to everyday conversations with chatbots. Analytic conversations often point to specific prior and require anaphoric references~\cite{Fast:Iris:2019}. Moreover, explicit statements may also be required to shift context (e.g., uttering ``retain,'' or ``start over.''). The prior WoZ study~\cite{tory:tc17:2019} only allowed participants to manage this context via utterances, but other approaches may offer more control without adding considerable overhead.


\vspace{1mm}
\noindent\textbf{DG4: Support Out-of-the-Box Visual Data Analysis.}\label{reduce-reliance}
LLMs often rely on a precise prompt in order to produce an appropriate response~\cite{Brown:LLM_fewshot:2020}. While analysts can explore ways to improve their prompts and utterances, for example via anaphoric references, 
end-users should be able to focus on conducting data analysis and not adhere to technology-specific vocabulary or need to repeatedly revise their utterances to produce desired results. 
Furthermore, effective prompt engineering requires specialized knowledge, and although recent efforts have aimed to help users explore automatically~\cite{Zhou:APE:2022} or interactively~\cite{Wu:LLM_chains:2022,Mishra:PromptAid:2023,Singh:ExplainingPE:2022} refine a prompt, these strategies interrupt the analysis and conversational flow.  An effective conversational agent should manage such considerations for the end user without requiring them to delve into the technical details.

\section{\sysname}\label{interface}

\begin{figure*}[ht]
  \includegraphics[width=\textwidth]{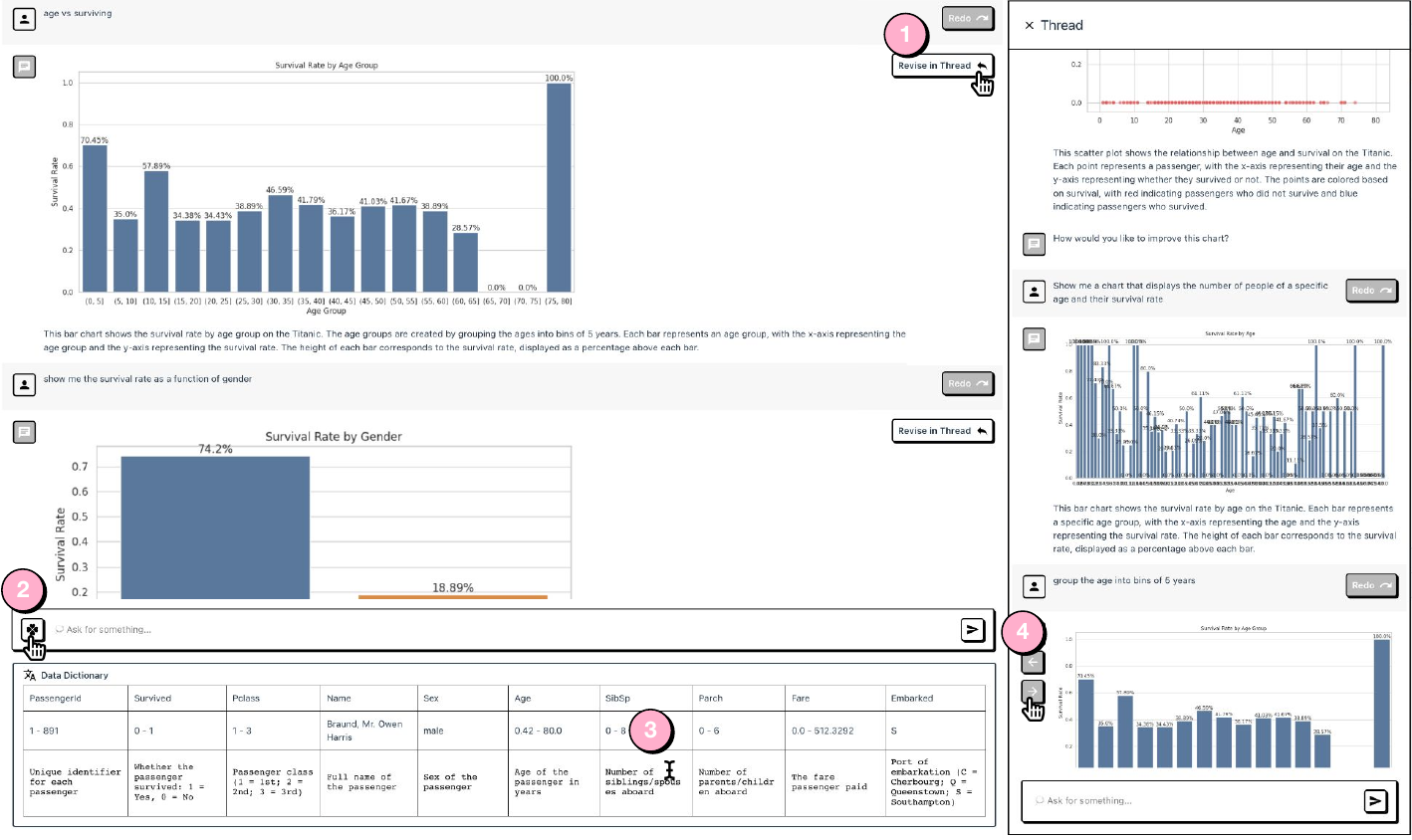}
 \caption{\textbf{Overview of~\sysname interface.} Users begin their analysis in the \mchat area (left), which shows the conversation history diarized by the user's and chatbot's responses. The chatbot's responses can be textual or data visualizations. To modify a visualization in collaboration with the chatbot the user can 1) open the \thread panel (right). The \thread panel retains a history of modifications that a user can return to and further refine at any time. Closing a \thread updates the visualization displayed in the \mchat conversation with the latest version in the \thread panel. When in the \mchat area, the user can 2) prompt the model to ``\textit{Show me something interesting}''. The \mchat also includes an 3) editable and persistent \ddict to help orient both the user and AI during visual data exploration. When model outputs are re-generated after clicking a redo button, the user can 4) click the arrow keys to revert to previous generative outputs. Unlike ADA, reverting to a prior response using the arrow keys does not delete utterances or responses that have already occurred later in the conversation.  This snapshot is from crowd-worker P13 (Section~\ref{usability}).}
 \vspace{-4mm}
  \Description{Caption}
  \label{fig:interface}
\end{figure*}

We developed~\sysname, a prototype system for conducting multi-threaded analytic conversations with an LLM chatbot.~\sysname allows users to upload tabular datasets, prompt the LLM with analytic questions, review visualizations, and refine them. We take design inspiration from prior V-NLI systems in addition to common messaging apps (e.g., Slack) and analytic notebooks, while also incorporating the design goals articulated in Section~\ref{design_goals}. 

\subsection{Overview}\label{interface:overview}
\sysname uses the GPT-3.5 model through OpenAI’s ChatCompletion API to generate visualizations and captions based on descriptions of the dataset and user utterances. The system comprises a web-based application frontend to collect utterances and display the chatbot's textual and visual responses as well as a backend to generate and execute code that produces the responses. Figure~\ref{fig:interface} provides an overview of the~\sysname interface, which comprises two components : (A) a~\mchat area and (B) a~\thread panel. The \mchat area serves as a conversational interface that also shows the trail of the user's analytic process (\textbf{DG1, DG2}). The \thread panel allows for refinement and modifications to visualizations in the \mchat. We use the separation between the \mchat and \thread to proactively manage the conversational context (\textbf{DG3}). We refined the user experiences while iterating over early designs of the prototype with feedback from initial pilot studies. 

We showcase~\sysname's capabilities primary through user studies (Sections~\ref{usability} and~\ref{user_interviews}) and usage scenarios (Section~\ref{use_case}). However, in  Appendix~\ref{app_capabilities} we also show examples of~\sysname executing different edit operations (Figure~\ref{fig:edit-ops}) and for different utterance intents (Figure~\ref{fig:intent_example}).

\vspace{-2mm}
\subsection{\mchat Area}\label{interface:main_chat}
The \mchat area includes utterance input area and retains an overview of the chat history. After the analysis of usability studies, we later included a~\lucky{} button which automatically prompts the LLM to \textit{``show me something interesting''}, an utterance entered by several participants.
The analysis history in the \mchat constitutes the diarized conversation between the chatbot and the end-user. The chatbot's response can be either text or static visualizations with accompanying explanatory captions. 

The chatbot's responses can be modified in two ways. First, if an error is produced or if the user is not satisfied with a visualization, they can click the \redo~button to automatically re-trigger another attempt (\textbf{DG2}). Both the current and prior responses are retained and the user can scroll through them using \arrowleft{}~or~\arrowright{} arrows. Unlike ADA, reverting to a prior response using the \arrowleft{} arrow does not delete utterances or responses that have already occurred later in the conversation, allowing the user to simply select which visualization should be part of the analytic context sent to the model (\textbf{DG3}). 
Second, a participant can actively modify a visualization by clicking on the ~\ithread{}~ button that triggers the \thread panel. Once a participant completes their refinements in the \thread, the visualization in the \mchat is also updated and becomes part of the main analytic context (\textbf{DG3}) without the intermediary refinement utterances to clutter it (\textbf{DG1}). 
As a last resort, users can also prompt the system to ``undo'' their previous prompt, which deletes the last user and assistant messages. This allows users to retype their utterances if their analysis has led to a dead end. 

Below the utterance input area is a \ddict~\cite{rashid:datadict:2020}, which is intended to help both the AI and the user understand the structure of their dataset (\textbf{DG4}).  The \ddict breaks down each attribute in the dataset according to its name, its value range or an example, and a brief description of the attribute. The attribute descriptions are automatically generated by~\sysname after uploading the dataset and are user-editable. The inclusion of each column's possible values and editable descriptions is intended to help the LLM  resolve potentially ambiguous attribute names as well as elicit higher-order reasoning about the dataset~(\textbf{DG4}; details follow in Section~\ref{system:data_sources}). 


\subsection{\thread Panel}\label{interface:thread}
The \thread panel comprises the same elements as the \mchat area (excluding the data dictionary). It is intended to support the user in refining analysis ideas and visualizations. There may be several reasons why a user may wish to initialize a thread. One that we have already described is the progressive elaboration that is common in analytic processes (\textbf{DG1}). Upon generating an initial visualization, the user may gain an insight that motivates them to further refine their utterance. This refinement can include modifying attribute bindings to different channels (by adding or removing them), modifying the data transformations (e.g., the bin width in a histogram), adding highlights or annotations, and adding reference or regression lines among other things~\cite{Kim:GraphScape:2017,ren:Chartaccent:2017}. The content of a \thread persists even after the participant closes the panel, allowing them to continue their analysis or share with others how their analysis evolved (\textbf{DG1}). 

Other reasons may arise from a misalignment between the user's intent or desired output and the response of the chatbot. For example, the chatbot may generate an encoding type that the user does not find effective (e.g., a pie chart when a bar chart is desirable). When utterances are ambiguous, the chatbot may also make incorrect choices of data transformations or visual encodings. The end-user can provide more specific details for the chatbot in follow-up utterances (\textbf{DG2}). \sysname is able to proactively accept and incorporate user feedback that corrects its outputs, we provide examples of these from user studies  and a concrete set of edit operations in Appendix~\ref{app_capabilities}-Figure~\ref{fig:edit-ops} .  
The use of the \thread panel also need not be limited to modifications. They can also be used to prompt the chatbot to explain the visualization it generated, provide suggestions for alternative visualizations, and extract insights from a visualization (for example, correlations, distributions, or trends~\cite{Srinivasan:Snowy:2021}).

\section{System Design}\label{sys_design}

\begin{figure}
    \centering
    \includegraphics[width=1\textwidth]{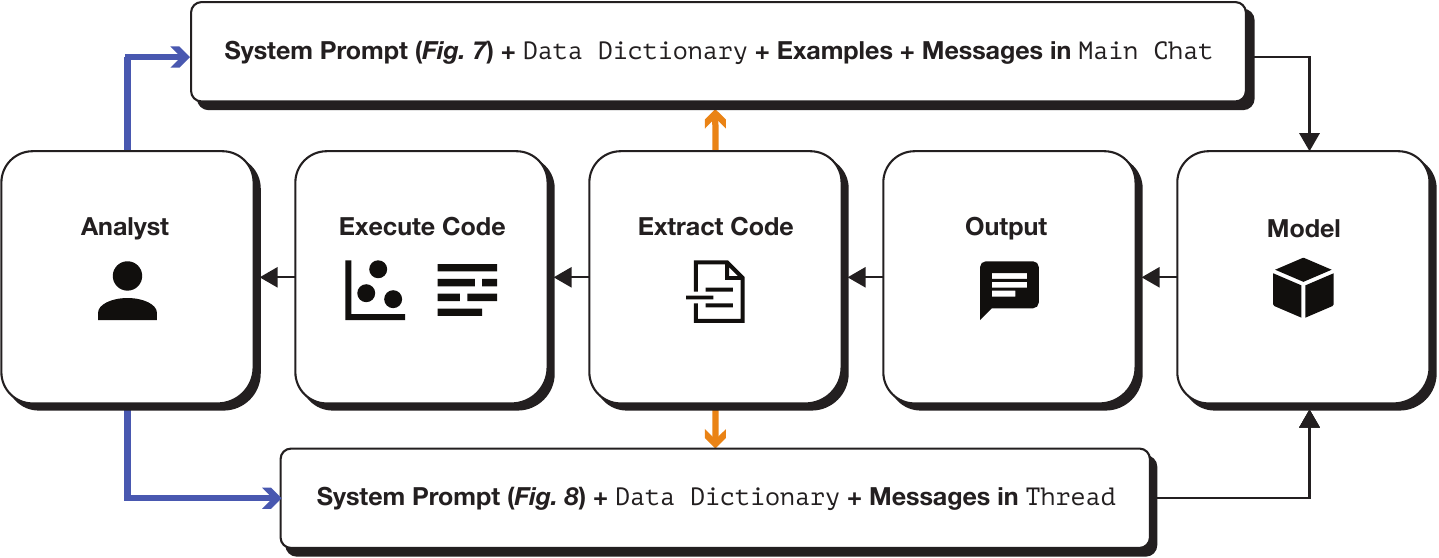}
    \caption{\textbf{Overview of~\sysname's code generation pipeline.}
    When an analyst enters a \textcolor{question}{\textbf{user utterance}}, the message is sent to GPT-3.5 with different system prompts and conversational context depending on whether they are interacting with the \mchat (top) or a \thread (bottom) panel. The corresponding system prompts are further described in proceeding figures. If the AI's \textcolor{response}{\textbf{assistant outputs}} contain executable code, the code becomes part of the subsequent conversational contexts. After the code executes, the user will see the rendered charts along with accompanying captions.
    }
    
    \label{fig:workflow}
\end{figure}

In this section we describe~\sysname's architecture that supports ingestion of user utterances and the display the result within the \mchat and \thread. We provide an overview of~\sysname's pipeline in Figure~\ref{fig:workflow}.

\subsection{Data Sources}\label{system:data_sources}

\begin{figure}
    \centering
    \includegraphics[width=0.4\textwidth]{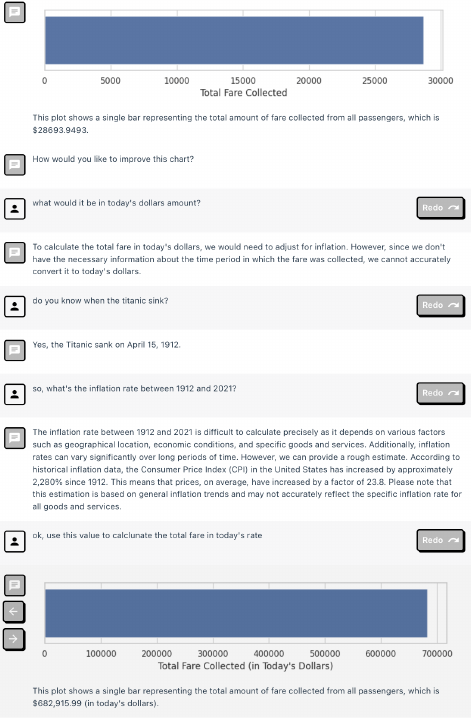}
    \caption{\textbf{Example of the chatbot obtaining and applying external knowledge to a analysis of the Titanic dataset in a \thread.} A user prompts the chatbot to compute the total fare paid and convert the amount to 2021 USD. The chatbot initially failed to produce a response, but with guidance from the user (Participant E02 in from the usability interviews (Section~\ref{user_interviews})) the chatbot eventually converted \$28,693.9493 to \$682,915.99. The initial fare sum is correct, but the amount adjusted in 2021 USD is incorrect;  the increase Consumer Price Index the chatbot used to compute this value is closer to a factor of 27.9 (according to the US Bureau of Labor Statistics), not the 23.8. However, the calculation ($28,693.9493 \times 23.8 = 682,915.99$) was performed correctly.}
    \label{fig:external_data}
\end{figure}

Participants can upload a single CSV dataset to carry out their data analysis. A string referring to the file's location on the local system is also stored and passed to GPT-3.5 as part of its code generation. When a dataset is uploaded, using the procedure outlined in Figure ~\ref{alg:dictionary-chain},~\sysname will automatically generate a \ddict which includes each column's name, data type, and a value range (for nominal or ordinal columns) or a non-null example value (for all other data types).
The LLM also generates a natural language description of each field.

The \ddict is used as part of future instructional system prompts to generate valid and executable code; it is also displayed to end users (Section~\ref{interface:main_chat}) as an editable reference table, where they can alter descriptions to provide more accurate information for the AI (\textbf{DG4}). These descriptions help the LLM identify and use the appropriate attribute and reason about its values (e.g., the utterances ``\textit{Filter gender by male}'' vs. ``\textit{show only male}''),  which supports a more naturalistic conversation out-of-the-box (\textbf{DG2}). We believe that descriptions about columns can also elicit higher-order reasoning from the LLM when necessary, for example whenever the user prompts the model to ``\textit{show me something interesting,}'' by leveraging knowledge about semantically related concepts captured within the LLMs embedding space. Future work presents interesting possibilities for extending this approach by augmenting the LLM with a knowledge graph~\cite{logan:KG:2019}.


\sysname can leverage knowledge that is external to the uploaded dataset as part of GPT-3.5's training data\footnote{At the time of this writing, the details of GPT-3.5's training data have not been publicly disclosed.} to augment their existing analysis. For example, with the Titanic dataset participants can ask~\textit{``Who was the captain on the Titanic?''}, which is not available in the dataset itself. Figure~\ref{fig:external_data} shows an example where a participant from the usability interviews (Section~\ref{user_interviews}) computes the total fare paid by participants and asks~\sysname to convert this value to US dollars in 2021.  ~\sysname produces an estimate with some guidance from the participant. To the best of our knowledge, such capabilities were not possible with prior V-NLI and conversational systems.

\subsection{Code Generation}\label{system:code_gen}

User utterances initiate the generation and execution of Python code to apply the appropriate set of data transformations using \texttt{pandas} and produce visualizations using \texttt{seaborn}~\cite{Waskom:seaborn:2021} and \texttt{matplotlib}~\cite{Hunter:matplotlib:2007}. In addition to producing a visual encoding, GPT-3.5 is also instructed to produce a text caption describing the encoding choices. Depending on how the utterance is posed,  GPT-3.5 sometimes also generates an interpretation of what is shown.

\begin{figure}[t]
\begin{center}
\shadowbox{%
  \begin{minipage}{0.93\linewidth}

Using a \texttt{pandas} script, \sysname creates a markdown table from the uploaded dataset that includes column names, data types, and data values structured as follows:
\vspace{1mm}

\begin{minipage}{1\linewidth}
\begin{lstlisting}
|            | Data Type  | Range or Example       |
|:-----------|:-----------|:-----------------------|
...
\end{lstlisting}
\end{minipage}
\vspace{1mm}

This markdown table and the original filename are used to produce the full \ddict using GPT-3.5.
\vspace{1mm}
\vspace{1mm}
Before the model is queried, we give it the following system prompt:
\vspace{1mm}

\begin{minipage}{0.02\linewidth}
\centering
\rotatebox{90}{\small{\texttt{System}}}
\end{minipage}%
\begin{minipage}{0.98\linewidth}
\begin{lstlisting}
We will generate markdown tables that begin with the following columns:
|            | Data Type  | Range or Example       | Description               |
|:-----------|:-----------|:-----------------------|:--------------------------| 
\end{lstlisting}
\end{minipage}

\vspace{1mm}

Finally, the following user message is sent to GPT-3.5:
\vspace{1mm}

\begin{minipage}{0.02\linewidth}
\centering
\rotatebox{90}{\small{\texttt{\textcolor{question}{User}}}}
\end{minipage}%
\begin{minipage}{0.98\linewidth}
\begin{lstlisting}
Here is a markdown table summarizing columns in [FILENAME]:
[PANDAS TABLE]
Generate descriptions for what the data column values mean and concatenate the descriptions to the markdown table.
\end{lstlisting}
\end{minipage}

  \end{minipage}%
}
\captionof{figure}{\textbf{Instruction prompt for generating a \ddict of an uploaded dataset.} }
\label{alg:dictionary-chain}
\end{center}
\end{figure}

\begin{figure}
\begin{center}
\shadowbox{%
  \begin{minipage}{0.93\linewidth}

In the \mchat, the following system instructions are provided to GPT-3.5:
\vspace{1mm}

\begin{minipage}{0.02\linewidth}
\centering
\rotatebox{90}{\small{\texttt{System}}}
\end{minipage}%
\begin{minipage}{0.98\linewidth}
\begin{lstlisting}
#1You are a data visualization assistant working with a dataset called [FILENAME]. 
#1You will answer questions with this dataset.
#2When you do not use the dataset, say "I can't answer this question with the data, 
#2but I found this:" You will use pandas and seaborn libraries to write Python 
#2code which prints a single PNG image in a bytes64 representation. You will then 
#2also print an alt-text caption for the chart.
#3Let's work responses out step by step to be sure you have the right answer.
#4The dataset: [DATASET DICTIONARY]
\end{lstlisting}
\end{minipage}
\vspace{1mm}

In the background, user conversations with the LLM are prepended with few-shot input-output examples as conversational context, e.g.:
\vspace{1mm}

\begin{minipage}{0.02\linewidth}
\centering
\rotatebox{90}{\small{\texttt{\textcolor{question}{User}}}}
\end{minipage}%
\begin{minipage}{0.98\linewidth}
\begin{lstlisting}
How many rows are in the dataset?
\end{lstlisting}
\end{minipage}

\begin{minipage}{0.02\linewidth}
\centering
\rotatebox{90}{\small{\texttt{\textcolor{response}{Assistant}}}}
\end{minipage}%
\begin{minipage}{0.98\linewidth}
\begin{lstlisting}[language=Python]
import pandas as pd
import seaborn as sns
import matplotlib.pyplot as plt
...

# Load the dataset
df = pd.read_csv('./workspace/{filename}')

# Get the number of rows in the DataFrame
num_rows = df.shape[0]

# Create the bar chart using seaborn, adjusting the size of the chart as needed
width = 8
height = 2
plt.figure(figsize=(width, height))
sns.barplot(x=[num_rows])
...

# Print a caption
print("This plot shows a single bar representing the total number of rows in the dataset, which is " + str(num_rows) + ".")
\end{lstlisting}
\end{minipage}
\vspace{1mm}

The example response is truncated; see the full examples in Appendix~\ref{app:prompt}.

  \end{minipage}%
}
\captionof{figure}{\textbf{Pre-defined context in the \mchat.} Although not exposed to the user, an exchange with the GPT-3.5 CodeCompletion API in the \mchat will include a system prompt and few-shot examples along with previous exchanges.}
\label{alg:main-chat}
\end{center}
\end{figure}

\subsubsection{Processing User Utterances}\label{process_utterances}
Prior V-NLI and conversational interfaces have used a variety of methods, from parse trees to heuristics, to decompose a participant's utterance, deduce its intent, and generate an appropriate output. We use GPT-3.5 to perform these tasks, providing it with instructional prompts to guide its visualization code generation or refinement. Subsequently, the relevant visualization code is extracted from GPT-3.5's original responses to display charts and captions to the analyst.

\subsubsection{System Instructional Prompts.} \label{system:inst_prompt}
The GPT-3.5 ChatCompletion API delineates three role types to manage multi-turn conversations: system, assistant, and user. System instructional prompts can define an LLM's \textbf{\textcolor[RGB]{117, 112, 179}{role}} and \textbf{\textcolor[RGB]{27, 158, 119}{desired behaviors}} and augment the user messages to achieve a desired response. 

In the \mchat, the full instructional prompt provided to GPT-3.5 contains the agent's \textbf{\textcolor[RGB]{117, 112, 179}{role}}, which is to answer questions about the dataset with visualizations, and \textbf{\textcolor[RGB]{27, 158, 119}{desired behaviors}}, which is to render and display a chart with an accompanying caption).
The system is also prompted to use \textcolor{purple}{\textbf{chain-of-thought (CoT) reasoning}}, which has been shown to improve the performance of LLM code generation~\cite{wei:chainofthought:2023}. The CoT prompt is reinforced with a set of few-shot examples (See Section \ref{few-shot}).
Lastly, the system message includes the previously generated \textbf{\textcolor{red}{data dictionary}}.
Crafting such instructional system prompts can require specialized knowledge to produce via a manual trial-and-error process \cite{liu2023pre}, and thus is something we do not expose to the user. We refined our prompting strategy over time and found it to qualitatively improve the chatbot's responses without the need for the user to input additional instructions (\textbf{DG4}).

In the \thread, the agent is given a different \textbf{\textcolor[RGB]{117, 112, 179}{role}}, which is to revise a previously generated visualization code to change visual encodings or make stylistic changes.
A user might also ask the model to present additional data columns, to aggregate the data in a new way, or to filter the data, which again requires information from the \textbf{\textcolor{red}{data dictionary}}.
However, the design of this instruction ensures that each user prompt is interpreted to be a question or a command that continues or modifies a specific analysis (\textbf{\textcolor{response}{the original code}}), as opposed to starting a new analysis that will require the synthesis of fundamentally different code.

\begin{figure}
\begin{center}
\shadowbox{%
  \begin{minipage}{0.93\linewidth}

\begin{minipage}{0.02\linewidth}
\centering
\rotatebox{90}{\small{\texttt{System}}}
\end{minipage}%
\begin{minipage}{0.98\linewidth}
    \begin{lstlisting}
#1Today we are being asked to revise the following code.
#2The revised code should print a new chart in a bytes64 PNG representation and 
#2print a new alt-text caption in the same way as the original code.
#5The code: [ORIGINAL CODE]
#4The dataset [FILENAME] contains: [DATASET DICTIONARY]
    \end{lstlisting}
\end{minipage}

  \end{minipage}%
}
\captionof{figure}{\textbf{Instructions in a \thread.} The role of GPT-3.5 in a \thread is to extend an analysis branch, using the \textcolor{response}{\textbf{original visualization}} as a reference for refinement or continuation.}
\label{alg:thread}
\end{center}
\end{figure}

\subsubsection{Few-Shot Examples.}\label{few-shot}
In the \mchat, the initial system message is followed by few-shot examples of pre-designed \textcolor{question}{\textbf{user}} utterances and the \textcolor{response}{\textbf{assistant's}} code responses (Figure~\ref{alg:main-chat}). These few-shot examples include how to transform data with \texttt{pandas} and use \texttt{seaborn} to achieve the desired image and text caption outputs. Utterances in the few-shot examples were: ``\textit{How many rows are in the dataset?}'' which teaches \sysname to display simple numeric responses as both a bar chart and within a text caption, and ``\textit{Plot the distribution of every continuous measure,}'' which signals how it should display visualizations with multiple facets. Both questions are also agnostic to the semantics of specific datasets. 
Appendix \ref{app:prompt} contains full details about the examples. 

\subsubsection{Code Extraction and Execution.}\label{extraction}
Responses from the GPT-3.5 ChatCompletion API are often not valid Python code excerpts that can be directly executed by~\sysname. 
Given that GPT-3.5 is a successor to both Codex and GPT-3 models, its responses may contain `friendly' textual responses (usually beginning with ``Sure!'' \cite{zou2023universal}) or explanations of a step-by-step approach in addition to code.
In order to prioritize analytic coherence (\textbf{DG1}),~\sysname only extracts a valid code block from model responses, which include the chart's caption.
When valid code blocks can be extracted, they are automatically executed and outputs (visualization and text) are pushed to the front-end interface. If valid code blocks are not generated,~\sysname responds by displaying the initial model response to the user.

For a variety of reasons, including API time-outs, the model may not generate executable code. In these cases, the participant can reattempt to generate the response from the front-end interface without the need to re-enter the utterance or provide a new one (Section~\ref{interface:main_chat}). 
In other cases, the model may also output a simple text response, for example returning the average fare paid as a number instead of producing a visualization. In some instances, the text outputs of~\sysname pose clarifying questions to the analyst, for example in responding to the utterance \textit{``tell me something that is cool''} an example response is \textit{``as an AI assistant, I don't have access to personal information or preferences unless shared with me. If you have any specific preferences or questions, please feel free to share them, and I'll do my best to assist you''.}

\subsubsection{Different Thread, Different Context}\label{context_mgmt}

\sysname provides different system prompts and conversational contexts to GPT-3.5 depending on whether the user is interacting with the \mchat and \thread panel (Figure~\ref{fig:workflow}). The conversational context is further tailored to a thread the user is inputting utterances to (Figure~\ref{alg:thread}). Code that is generated within a thread is appended as prior context to subsequent utterances. Alternating between  \thread and \mchat allows the user can proactively manage context (\textbf{DG3}). 
This is because the \thread panel contains only the context relevant to that particular visualization and analysis while the \mchat will only include the utterance and results of charts as refined in each thread.

\subsubsection{Alternative Visualization Libraries} We explored other visualization libraries but found that they resulted in less reliable responses. In some instances, the libraries were too new to produce reliably executable code, as was the case for \texttt{altair}~\cite{vanderplas2018altair}. We also explored different languages for visualization specifications. Prior work \cite{Poesia:Synchromesh:2022,Dibia:LIDA:3034} has explored the use of JSON-formatted \texttt{Vega-Lite} specifications to produce visualizations using LLMs, but we did not use these approaches for two reasons. First, 
using \texttt{seaborn} allows for direct integration with its dependencies \texttt{statsmodels}, \texttt{scipy}, and \texttt{fastcluster} for advanced data modeling. Keeping code for the data operations and visualization in the same ecosystem improved~\sysname's ability to reliably generate executable code. Second, one advantage of using \texttt{seaborn}~\cite{Waskom:seaborn:2021} or \texttt{matplotlib} was that GPT-3.5 would produce comments to document its data transformations and visualization processes, which is a useful exposition of its chain-of-thought reasoning~\cite{Drori2022} that we could leverage in the future. However, getting equivalent information from a JSON specification proved to be difficult. 





\section{Usability Survey}\label{usability}
We conducted a survey of~\sysname's usability with 40 participants. The goals of our usability survey were to establish the diversity of use for~\sysname and participants' self-reported perceptions of their sessions, including the quality of responses and visualizations.

\subsection{Session Protocol}\label{usability_session}
All participants completed an initial intake that gathered additional information about their demographics, their programming, and data analysis background. Participants were then shown a video approximately two minutes in length\footnote{This video can be viewed here: \url{https://www.youtube.com/watch?v=Ll0nPtD3kMc}} providing an overview of~\sysname~, including examples of different queries and the use of threads. We use the Titanic dataset from~\cite{tory:tc17:2019} for this study. Participants were then directed to a the~\sysname~and prompted to conduct an analysis that identified \textit{Which factors increased one's chances of surviving the Titanic shipwreck?} 
We did not restrict participants from simply inputting this question to~\sysname. 
Participants received a completion code only after they spent at least three minutes conducting the analysis and after creating at least one thread. Following their session, participants were guided to a set of questions that were used as an attention check to establish whether or not they meaningfully explored the dataset. Finally, participants were asked to rate their session, chatbot responses, and visualization results on a series of five-point Likert scale questions while also providing justifications for their responses in free-form text.

\subsection{Data Collection and Analysis}\label{usability_data}
In addition to each participant's session log, we collected participant data from a survey that gathered Likert scale and free-form textual responses. We summarize all Likert scale questions using simple descriptive statistics. Participant utterances were analyzed using a ground theory approach. The session logs include a full transcript of the conversation, which constituted the participants' typed utterances as well as the system and AI Agent responses, including both the code generated for data transformations and visualization specifications and finally a PNG image of the resulting visualization presented to the user. During our analysis, we replayed these user sessions using ~\sysname~ in read-only mode.

\subsection{Participant Recruitment}
Crowd-sourcing platforms are increasingly used by visualization researchers to recruit participants for a variety of tasks, including, as we do here, evaluating user interfaces~\cite{Borgo:VisCrowdsource:2018}. We recruited participants from the Prolific platform\footnote{\url{https://app.prolific.co/}}, aiming to collect a balanced gender sampling. 
While our study does not require specialized analysis skills or knowledge of the Titanic dataset, our inclusion criteria aimed to increase participants' likelihood of having some analytic experience (for example, by having knowledge of certain programming languages or systems). We also restricted our study to English speakers in the U.S., with at least 100 approved tasks. We paid a rate of \$3.50 for 20 minutes of effort, which equates to \$10.50\textbackslash hour---higher than the US federal minimum wage. We screened all participants' textual responses and session logs, removing participants that did not use \sysname in a manner relevant to the study tasks.

\subsection{Findings}

\subsubsection{Participant  \& Session Overview} Our study had a roughly even breakdown of participants that identified as female (n=19, 47.5\%) or male (n=21; 52.5\%). A total of 7 (17.5\%) participants indicated they were currently students. Nearly half (n=17; 42.4\%) of participants had background in one of either computer science, data science, machine learning, or natural language processing. Moreover, 35 (70\%) reported routinely conducting data analysis, primarily with spreadsheet programs (Excel or Google Sheets; n=26; 65\%) and/or scripting languages (Python or R; n=9; 22.5\%); two participants reported having previously analyzed the Titanic dataset. Despite the technology's recency, 9 participants (22.5\%) reported having previously used an LLM tool like ChatGPT to carry out data analysis. 

The average session was approximately 24 minutes. The shortest session was 10 minutes and the longest was 65 minutes; the majority of sessions were closer to the average time than either of these extremes. Although there was nothing restricting what prompts participants could enter, it was interesting to note that approximately half began their session simply by pasting the task prompt into~\sysname. The rest began their session by asking other questions.

\subsubsection{Survey Responses}

\begin{figure*}[t!]
    \centering
    \includegraphics[width=\textwidth]{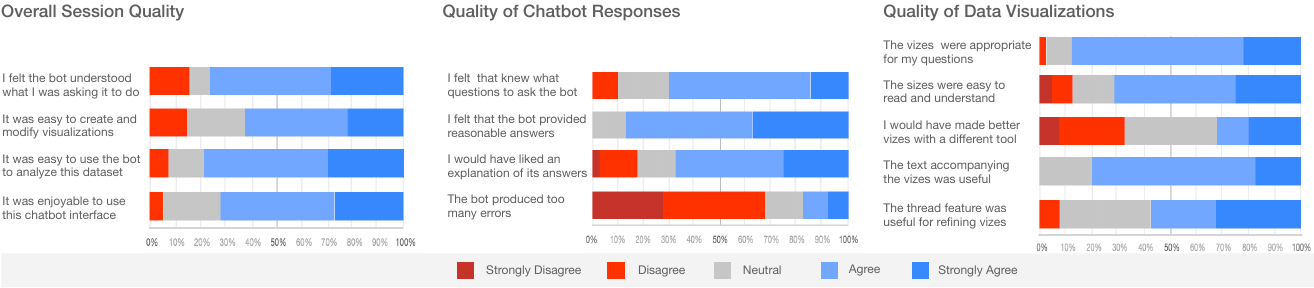}
    \caption{\textbf{Summary of participants impressions of interacting with~\sysname}.}
    \label{fig:likert_overall}
\end{figure*}

Participants perception of the user interface, chatbot's responses, and the quality of the visualizations are summarized in Figure~\ref{fig:likert_overall}. Overall, the majority of participants agreed or strongly agreed that it was easy (77.5\%) and enjoyable to use the Chatbot (72.5\%). The majority also felt that the Chatbot understood what they were asking it (77.0\%) and that the visualizations it produced were appropriate for the utterances they inputted (87.5\%) and were easy to understand (71.8\%). While some participants encountered errors while using the Chatbot, the majority (67.5\%) did not find these errors to be onerous. While in aggregate participants responses report an overall positive experience, individual sessions still varied. For example, P11 reported that~\textit{``the answers seemed to be well thought out and professional. It seemed to know the exact answer I was looking for while providing additional information.''} While P39 took the opposite view and found that~\textit{``it felt rather rough, like something still being worked on. It took a bit of effort to get the data I wanted.''} From participants textual responses that justified their ratings, we identify several recurring themes. 

\vspace{1mm}
\noindent{\textbf{It was sometimes necessary to modify an utterance to achieve the correct response.}} Several (n=6) participants indicated that the chatbot did immediately understand their utterance and they \textit{`` had to word things differently sometimes to get the results I was looking for''} (P26). 
At times the Chatbot interpreted an utterance too literally as P05 observed:~\textit{``the Chatbot will generally give you pretty much what you asked for, so it is important to understand how to talk to the Chatbot''} P06 noted that ambiguity of the utterance contributed to the importance of phrasing: \textit{``I noticed that I had to word my questions carefully and specifically. For example, I asked the bot 'Which factors increased one's chances of surviving the Titanic shipwreck?', a very general and broad question, and the chart it came out with was basically empty. This was understandable considering the variety of factors that could be given as an answer to the question''.} 
This observation is in conflict with \textbf{DG2}, but the majority of participants did not report having this experience. Overall this observation demonstrates how variable interactions with  Chatbots can be.

\vspace{1mm}
\noindent{\textbf{Participants thought visualizations were appropriate for their questions, but needed some improvements.}} P18 expressed the general sentiment that ~\textit{``visualizations were clear and a majority of the time were what I asked''}. Again, there was a spectrum of experience with P20 reporting that~\textit{``The visualizations were just okay. They often did not correlate to what I asked for, were not labeled well, and were not visually appealing''} 
Participants also had different experiences modifying visualizations.  One participant appreciated  that~\textit{``modifying them [the visualizations] only required simple commands.''} (P25), while others reported that sometimes~\textit{``it [the chatbot] also pulled from incorrect data and had a hard time fixing it''} (P30). One participant provides a specific challenge: \textit{``creating a chart breaking down survival rate by age, since it tried to do a bar graph for each individual age rather than breaking them into larger age ranges.''} 

\begin{figure}
    \centering
    \includegraphics{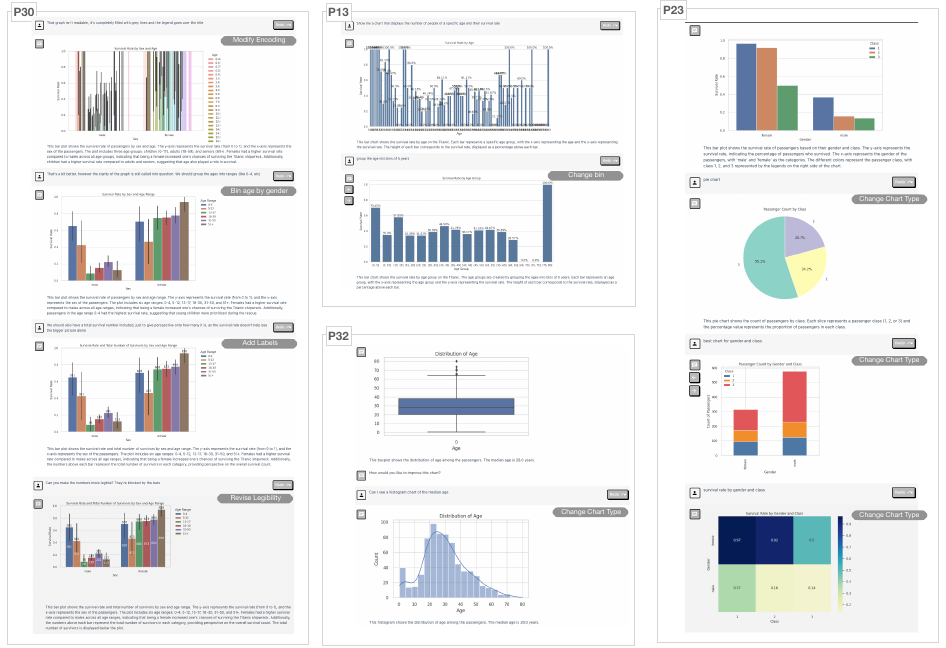}
    \caption{\textbf{Example use of threads to modify analyses and charts.} We added grey pills to summarize the key refinements.}
    \label{fig:prolific_chartmod}
\end{figure}

\begin{figure}
    \centering
    \includegraphics{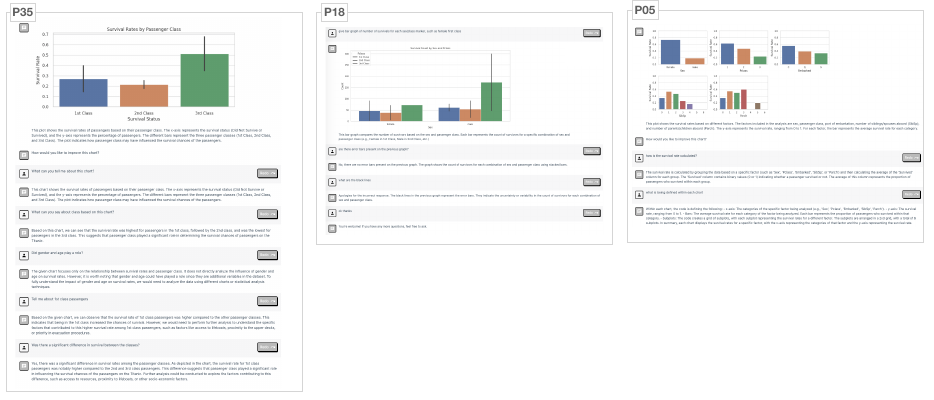}
    \caption{\textbf{Example use of threads to explain information in the chart}}
    \label{fig:prolific_explain}
\end{figure}

\vspace{1mm}
\noindent{\textbf{Threads were used to modify visualizations \& ask for more information.}} The majority (57.5\%) of participants thought the threads features was useful, and from reviewing the sessions logs we observed its most common use was changing the encoding types. However, the overall number of ways participants used threads was quite varied. In Figure~\ref{fig:prolific_chartmod} we show examples of how participants modified visualizations through the useful of threads. P13 was enthusiastic about their experience:~\textit{``I really enjoyed the thread feature to edit the graphs and make them a bit easier with more information on what was wanted''}. P31 describes their analysis process:~\textit{``
When asking for the median age, it produced a box plot that was difficult to analyze. I opened threads and asked it to change it a histogram, which was easier to read.''} Participants also used  threads to get explanations about the chart (Figure~\ref{fig:prolific_explain}). The explanations were useful to P05 who found that~\textit{``the chatbot helped further clear things up  [about the chart] when I asked about it in the threads''} P18 had an interesting scenario of correcting the chatbot when it claimed that error bars were not present in the chart. However, it was evident that the chatbot was not relying entirely on the chart and that data when creating explanations. In the P35's explanatory context, the chatbot provides an incorrect explanation of the chart; it hallucinated the response. We observed other instances of this kind of hallucination occurring, but it was mixed in with relevant and correctly generated responses -- making it difficult to precisely identify factors that caused this to occur. 

Overall, from this usability survey the use of threads was generally shallow. We noted that participants appearing to have stronger analytic backgrounds made more extensive and meaningful uses of threads, and were more likely to report positive experiences using them. This observation was the motivation for our follow-up study interviewing analytic experts. 

\subsection{Summary and Implications}
In general, participants experiences align with our own investigation of the data from Tory and Setlur~\cite{tory:tc17:2019} and showcase the diversity of experience toward AI Chatbot use for data analysis. While individual sessions showcase a spectrum of experiences, the majority of participants reported being both pleasantly surprised and delighted by the accuracy of the chatbot's interpretation of their utterances and its responses. We used participant's responses to revise~\sysname by addressing errors these sessions surfaced and improving the overall responsiveness of the bot.

\section{Usability Interviews}\label{user_interviews}
We follow up on the results of the usability survey by conducting semi-structured interviews with 10 individuals with data and visual analysis expertise.

\subsection{Study Protocol}\label{interview_protocol}
The participants completed an initial intake identical to the usability survey (See Section~\ref{usability}). Following this intake, participants were contacted to participate in a 30-minute follow-up session to review their analysis. In some instances, participants did not complete their analysis before the follow-up session, when this was the case they completed their analysis during the follow-up session, extending the total session time up to 45 minutes. Each session included one study administrator and a note-taker. The study administrator primarily followed and script to ask participants about their experience; where appropriate the administrator also followed up with improvised probing questions. 

\subsection{Participants}\label{interview_participants}
We set a recruitment goal of ten participants with data and visual analysis experience through social media. Participants who completed the intake form but that were ineligible to participate in this study, primarily because they did not have the requisite background, were not contacted for the follow-up study and their results were not used.  We continued to leave the intake form open until ten suitable participants were identified. The participants we predominately male (n=9) with technical roles as a data scientist/engineer (n=2), data visualization experts (n=4), business analyst (n=4), and machine learning engineer (n=1); participants could select more than one role. Three participants also identified with other roles including engineer managers (n=2) and one technical writer. Participants reported using Python/R (n=2), Tableau/Lookr/PowerBI (n=4), and Excel/Google Sheets(n=4) to conduct data analysis. Participants also reported using Dataiku (n=1), Spotfire (n=1), and Smartsheet(n=1). Three participants said they had used LLMs to conduct data analysis. Two participants had previously used OpenAI's ADA to conduct analyses of different datasets.

\subsection{Data Collection \& Analysis}\label{interview_data}
As with the previous usability survey, all participants rated their experiences using Likert Scale data; participants did not have to provide textual justification for their scores as this information was collected during the follow-up interview. We again collected the full session logs for all participants that we reviewed with them during the follow-on session. Finally, all sessions were recorded and transcribed. We conducted a thematic analysis\cite{braun2006using} to summarize participants' impressions of the tool's usability as well as questions related to their analysis, including their analytic intent behind prompts and the quality of AI's responses to their prompts. Themes and subthemes were synthesized using a ground theory open-coding approach~\cite{charmaz_constructing_2006}. 


\subsection{Findings}\label{interview_findings}


\begin{table}
  \caption{Categories of User Feedback}
  \centering
  \small
  \begin{tabular}{>{\raggedright\arraybackslash}p{2.5cm} >{\raggedright\arraybackslash}p{3.5cm} p{8cm}} 
  \toprule
  \textbf{Theme} & \textbf{Sub-themes} & \textbf{Example Quote} \\
  \midrule
  \multirow{2}{2.5cm}{Role of~\sysname in Analysis ($n=8$)} 
  & Sociotechnical ($n=8$) & ``It's good to get familiar with a dataset and potentially be able to share data visualizations to their team or to somebody higher up'' [E06] \\
  \cmidrule(l){2-3}
  & Data Engineering ($n=7$) & ``How do I kind of define these variables that I want to work with... and know that my chart is using that?'' [E01] \\
  \midrule
  \multirow{2}{2.5cm}{Reflections on System Features ($n=10$)} 
  & Ability to Organize Analysis ($n=10$) & ``Keeping the context of the conversation intact is actually a very useful feature... you can walk down this garden path but still not lose your train of thought.'' [E03] \\
  \cmidrule(l){2-3}
  & Having a Two-Way Conversation with AI ($n=9$) & ``I kind of learned I guess that maybe for more complex things you have to be a little more specific.''[E08] \\
  \cmidrule(l){2-3}
  & Enchantment and Captivation ($n=7$) & ``I was surprised to get graphs... It was a nice surprise to actually get those.'' [E05] \\
  \bottomrule
  \end{tabular}
  \label{tab:themes}
\end{table}

We summarize participant impressions and suggestions in Table \ref{tab:themes}, which shows the themes and sub-themes that emerged from our thematic analysis. We identified two primary themes centering on 1) the role of~\sysname in data analysis and 2) reflections on existing features and suggestions for additional features. Participants reported an overall positive experience with only one reporting  having an overall negative experience. This participant (E01) expressed wanting greater control of~\sysname's outputs and felt that natural language inputs did not enable such fine-grained refinements. Despite this, they still report having  ~\textit{''a magical moment [...]''} using~\sysname \textit{``that's getting into Star Trek Communicator Land [because] I don't have to say, rebuild this chart or whatever. It just got it.''}. 



\subsubsection{The Role of \sysname in Data Analysis.}
Participants' overall impressions of~\sysname's usability largely align with the results previously reported in Section~\ref{usability}. From the interviews, we were able to get a better understanding of participant's overall analytics workflow and the ways the conversational agents fit into them. Two themes consistently emerged. First, nearly all (n=8) participants commented on the ability to share the analysis with others. Second, participants wanted to explore know the capabilities of~\sysname on larger and streaming datasets that were reflective of their common workflows. We expand on these themes here:

\vspace{1mm}
\noindent\textbf{Sociotechnical Considerations.} \ system's role was perceived predominantly as a useful tool for rapid exploratory analysis of a dataset.
However, exploratory analysis exists as part of a socially structured data science workflow, and participants were quick to speculate how the data artifacts generated by \sysname could be used to facilitate collaboration.
As E06 noted, \sysname was \textit{``good to get familiar with a dataset''} with the objective of \textit{``potentially be able to share data visualizations to their team or to somebody higher up.''}
One participant (E02) speculated that in the near future, LLMs would be integrated into messaging applications such as Slack or Microsoft Teams where multiple analysts could collaboratively analyze a dataset.

On the other hand, two participants (E08 and E09) envisioned \sysname as a separate tool that may serve as a replacement for Jupyter Notebooks, and the figures could be exported and shared with others.
With these considerations in mind, it was natural for E04 to question \textit{``will it take into account if it’s for leadership, or in tech, or a data scientist?''} if they could control the persona of the LLM.

\vspace{1mm}
\noindent\textbf{Data and Processing.} Participants also considered what would be required before using \sysname for data analysis.
For example, E07 wondered\textit{ ``How will it handle large datasets''}especially when the tables were in wide format (e.g., each column is a date).
Particular considerations to the data engineering workflow were given when participants were used to making dashboards for tracking assimilated and streaming data.
Even if the data were tidy and stable, E01 wondered\textit{ ``how do I kind of define these variables that I want to work with... and know that my chart is using that?''}
Although in E01's session, they were happy that \sysname was able to automatically infer that they wanted to plot survival rate (instead of survivor count) without explicitly specifying the calculation, they were still concerned about the lack of provenance tracking with regards to the AI's data transformations.


\subsubsection{Reflections on System Features}
While using~\sysname participants also reflected on ways to expand its capabilities to further support them in data analysis. We break these into three sub themes:

\vspace{1mm}
\noindent\textbf{Ability to Organize Analysis.} Expert participants were overwhelming positive about the ability to organize their analysis using \sysname, particularly with the threads feature.
E07 noted that they wouldn't \textit{``want this information in the thread to be used in the main chat."} and E08 reflected on how they effectively used the feature as they \textit{``Switched to main chat when thread was different enough to start over."}
The conversation in the \mchat represented where the major data artefacts should be stored, and E03 noted that \textit{``keeping the context of the conversation intact is actually a very useful feature... you can walk down this garden path but still not lose your train of thought''}  \textbf{\textit{Collectively, these findings mean that \sysname could be effectively used to keep track of the analysis process to maintain a coherent conversational memory not just for the AI, but also for the user.}}


On the other hand, several participants noted that additional features could support the organization of data artifacts, such as bookmarking messages in both the \mchat and the \thread.
The ability to title Threads was also a frequently requested feature to support data sensemaking. E10 expressed that while the threads feature was useful for follow-up on analyses, but felt that it could still be limited as the analysis grew more complex.  Moreover, they also pointed to issues with generative AI tools when~\textit{``trying to go back to a specific place in the conversation, it's not obvious where the entry points are to each thread.''}. They further expanded that when analyzing data they \textit{``got this exploding fractal mess, because that's what happens inside. And then ultimately, I'll go back and try and pick out the ones that are liked.''} However, they acknowledge this was a common problem with all analysis tools and that extending threads is a promising way to address these issues. While E08 also speculated about a multi-level thread feature, they noted their concerns with \textit{``going too far with threads and subthreads... it can be something like Slack, but something like Reddit would be too much.'' }

\vspace{1mm}
\noindent\textbf{Having a Two-Way Conversation with AI.} Most of the negative feedback about \sysname concerned the passive role the AI seemed to play in the design of charts.
E01 acknowledged that in order to analyze data effectively, people need to have expertise in both data analysis and the dataset itself. Before entering their first prompt, E09 took a significant pause wondering what questions they should ask even though the dataset was about an infamous historical event with well-known narratives. Starting with AI suggestions would have been useful for them to get into the flow of analysis immediately.

Analysts would rather be analysts and not prompt engineers.
Four out of nine expert users hoped to delegate aspects of the prompt design itself to the AI in the form of AI-drive chart recommendations prior to the start of data analysis.
Such recommendations could take the form of a ``suggestion box" similar to Show Me \cite{Mackinlay:ShowMe:2007}.
Experts will still have a hard time recollecting the names of charts (e.g., 100\% stacked bar charts) and E08 mentioned that a chatbot making suggestions like \textit{``Would you like to see a chart that combines [these two variables]?''} would be useful.
Without such suggestions, as reflected in the behavior and utterances of 7 out of 10 participants, LLM users are \textit{``mostly just playing around with what [the AI is] capable of''} at the beginning of sessions. To this end, E08 also noted that in-depth explanations of why AI generated its responses or threw unresolvable errors would be useful for users to \textit{``form a mental model'}' for how to prompt the AI model more effectively. 

Both E04 and E05 noted that the model's text responses were ``\textit{very good}.'' However, E05 speculated that an information bubble that reveals some out-of-the-box supporting narratives about why a generated visualization might be important in the data context may be useful to have.

\vspace{1mm}
\noindent\textbf{Enchantment and Captivation.}
Generally, users were pleasantly surprised by \sysname's ability to generate shareable visualizations from natural language alone.
When comparing to existing public-facing AI systems, E05 was pleasantly surprised that \sysname generated charts by default, remarking that generally \textit{``graphs are a good way to''} get \textit{``an answer as quickly as I could.''} Concurrent with our analysis of ADA (Section~\ref{formative_study}) one participant stated that when they~\textit{``use ChatGPT and other similar tools and it's mostly text. Unless I specifically say: "Present this in the table," or "Give me steps," or something like that,'' }. This highlights the importance of LLM-driven visualization interfaces to go beyond the ability of foundation models by automatically rendering code into desirable images, especially because visual data exploration is usually what is expected in earlier phases of data analysis.
E02 remarked they saw interactive systems like \sysname \textit{``as the future of analytics.''}

Whenever~\sysname was able to correctly assume design choices without explicit prompt engineering, users were captivated by the result.
E01 noted that it was \textit{``magical''} that they did not need to specify the number of bins in a histogram.
They shared another such experience: \textit{``When I did the survival by pclass [...] what I really meant was the rate, then it gave me exactly the rate. Those are perfect to me.''}
When E03 received a response to an utterance even though they did not specify a visual encoding, they remarked that \textit{``In my mind I would have never gone with a stacked bar chart as my first option''} but that it was the appropriate response upon reflection.
E07 noted that such automation highlighted the potential for NL-driven visual data analysis to empower them to get \textit{``a lot of analysis done.'' }
E01 speculated that AI could eliminate much of the time and effort traditionally needed to make complex visualizations and highlighted the potential for LLMs to overcome the traditional trade-offs between learnability and expressibility.

\section{Use Case}\label{use_case}
\begin{figure}[b]
    \centering
    \includegraphics[width=1\textwidth]{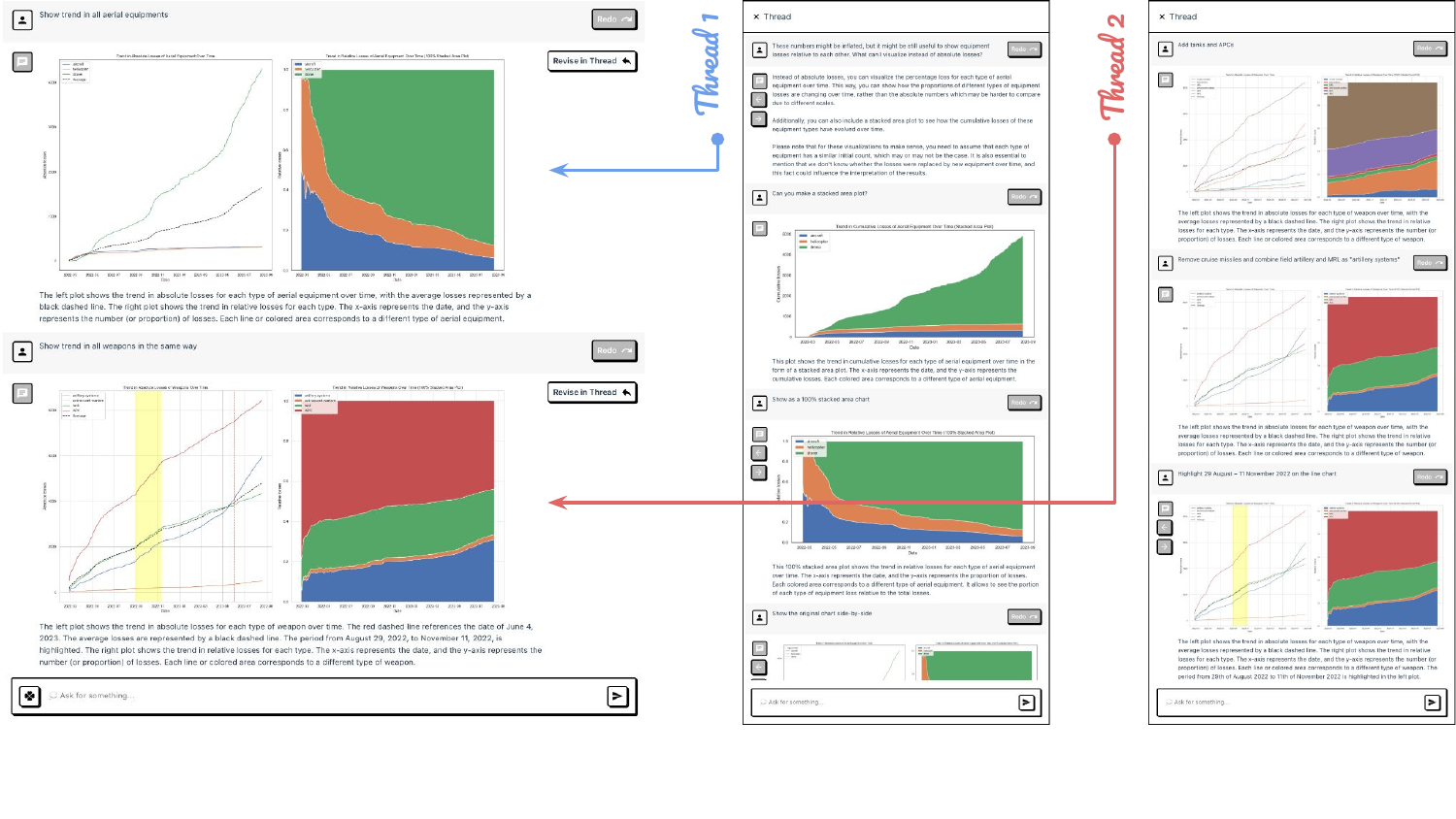}
    \vspace{-15mm}
    \caption{\textbf{Example analysis for Ukraine-Russian war dataset}. Two visualizations and shown in the \mchat area in addition to their modifications via two separate \thread panels.}
    \label{fig:demo}
\end{figure}

Finally, we demonstrate the use of \sysname on a dataset collected after GPT-3.5's knowledge-cutoff date of September 2021.
The Armed Forces of Ukraine has published data related to its military operations since the first week of the war, which began in 2022.
A structured dataset was compiled from the Ukranian government's website as daily reports and continues to also be updated weekly on Kaggle.
The dataset contains information about the cumulative number of losses per day sustained by Russian forces, including both military personnel and equipment. We consider a usage scenario of a data journalist, Sam, as they analyze this data using~\sysname. A summary of Sam's analysis is shown in Figure \ref{fig:demo}.

\begin{figure}[t]
    \centering
    \includegraphics[width=1\textwidth]{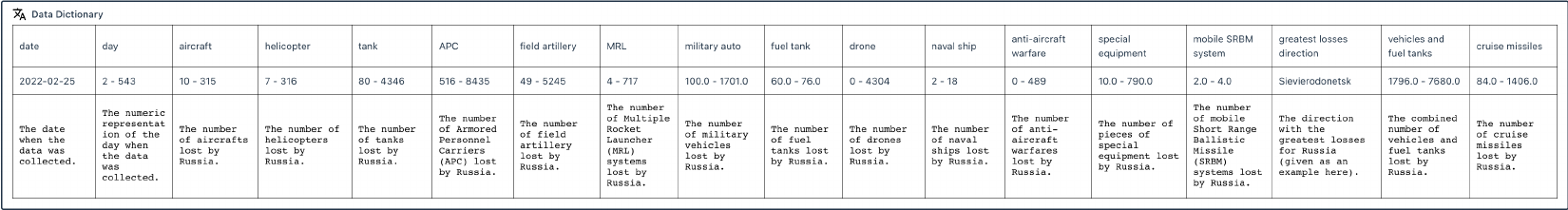}
    \caption{\textbf{Ukraine-Russian war dataset}. The \ddict for this dataset was automatically derived by~\sysname.}
    \label{fig:demo-dictionary}
\end{figure}

Preparing for a story pitch meeting,  Sam wants to quickly explore the dataset 
to find compelling visualizations that can be used to support their investigation.
They upload the dataset and
\sysname generates a data dictionary including the description of each column (Figure \ref{fig:demo-dictionary}).
Sam begins by typing: \textit{``Show trend in each equipment.''}
\sysname correctly responds with a line chart displaying the cumulative loss of each Russian piece of equipment throughout the war.
However, all 15 equipment types are difficult to distinguish using a single color palette.
Sam directs~\sysname to `\textit{`Show trend in all aerial equipments'' .}
The simplified line chart now clearly communicates that
Russian drone losses are rising day by day.
This is in stark contrast to the stagnation in the cumulative losses of aircraft and helicopters, even though the war has escalated.

Sam understands that
since the data is maintained by the Ukrainian forces, it may inflate the number of true Russian equipment losses.
So Sam opens a \thread next to the line chart and asks: \textit{``These numbers might be inflated, but it might be still useful to show equipment losses relative to each other. What can I visualize instead of absolute losses?''}.
\sysname responds by recommending a stacked area plot to present the data, so Sam further elaborates by first uttering \textit{``Can you make a stacked area plot?''}  and 
subsequently to \textit{``Show as a 100\% stacked area chart''} and finally
to \textit{``Show the original chart side-by-side''} -- the result is combining line and area charts can highlight both absolute and proportional values.
Finally, the line chart would seem more insightful if it depicted the average loss across all equipment types as a baseline to compare the trends of each piece of equipment.
Sam directs~\sysname to draw the average loss as a dashed line on the line chart and exits the \thread.

Sam is pleased with their first visualization, which they see now in the \mchat with the AI.
However, the visualization only depicts a small subset of the data.
Interested in building a narrative about well-known Ukrainian counteroffensives,
Sam directs~\sysname to \textit{``Show trend in all weapons in the same way'' }(i.e., the same way as the previous result which Sam heavily refined in a separate \thread).
Although \sysname's new visualization correctly presents some weapon losses as a paired line and area chart, Sam is not entirely happy with the weapons shown.

Sam opens a \thread next to the figure.
They direct~\sysname to \textit{``Add tanks and APCs''} since tanks and APCs (Armored Personnel Carriers) are also used in offensive operations.
Sam then directs the AI to \textit{``Remove cruise missiles and combine field artillery and MRL as artillery systems.'' }
Cruise missiles are single-use ammunition, unlike the rest which are multi-use equipment;
field artillery and MRLs are both mid-range artillery systems.
Sam now notices visual trends emerging in the line chart: two notable periods where there were relatively steeper inclines. They wonder if they correlate with the launch of major counteroffensives.
Sam directs~\sysname to \textit{``Highlight 29 August – 11 November 2022 on the line chart''} the duration of Ukraine's first counteroffensive, and is pleased to see the first dramatic increase across Russian weapon losses overlaying perfectly with the highlighted period.

Finally, the second counteroffensive was launched on the 4th of June, 2023 \footnote{https://www.reuters.com/world/europe/russia-says-its-forces-thwarted-major-ukrainian-offensive-2023-06-04/}.
Sam directs~\sysname to \textit{``Draw a reference line on June 4, 2023''} and notices that the dramatic rise in Russian artillery systems lost corresponds reasonably with the date.
Sam exits the \thread and is ready to share the \mchat, complete with charts plus titles and captions explaining their visual elements, with their colleagues in the pitch meeting.

\section{Discussion}\label{discussion}

The capabilities of Large Language Models (LLMs) for generating code and executing data analysis tasks have introduced a new paradigm, potentially overshadowing existing techniques for Visual Natural Language Interfaces (V-NLIs). However, it is vital to acknowledge the limitations of LLMs while appreciating their potential. In our exploration of these capabilities, we revisited the findings of a previous Wizard-of-Oz study~\cite{tory:tc17:2019}. Our analysis revealed that while existing LLM conversational agents excel at generating individual visualizations, they face challenges in accommodating shifts in conversational contexts, especially when transitioning from refining previous analyses to introducing new queries. We also noticed an intricate issue where the iterative refinement inherent to data analysis complicated the conversational history, making it difficult to navigate and comprehend

To address these challenges, we introduced a system, \sysname, designed to harness the potential of LLMs as analytic agents. \sysname facilitates multi-threaded conversations, enhancing a chatbot's contextual focus and improving conversational coherence. Through comprehensive user studies – one with crowd workers and another with analytic experts – we ascertained that \sysname can adeptly handle a vast range of participant utterances, ensuring smooth transitions between threaded and non-threaded dialogues. Notably, \sysname proved its competence even with datasets beyond the knowledge scope of LLMs. Collectively, our research underscores both the strengths and weaknesses of LLMs in their role as analytic agents. In the following sections, we delve into the limitations of our study and chart out potential avenues for future research.

\subsection{Limitations}
Our research has several limitations. \textit{First, a direct comparison with the closed-source ADA system remains challenging due to the undisclosed details of its implementation and update cadence. }Nonetheless, our open-source release of \sysname offers a platform for the visualization and HCI communities to iterate and potentially match or surpass ADA's capabilities. Over time, the research community may develop a wider suite of open-source tools that not only emulate ADA but also pave the way for more transparent comparisons. \textit{Second, the absence of quantitative benchmarks for visual analytic conversations led us to rely primarily on qualitative evaluations based on conversational shifts from~\cite{tory:tc17:2019}}. While the lack of quantitative benchmarks limits our assessment of \sysname's performance at scale, our approach offers invaluable insights into real-world analytic chatbot interactions. Future work could involve the establishment of robust benchmarks to assess the efficacy of such systems. \textit{Third, we intentionally chose the Titanic dataset to maintain alignment with~\cite{tory:tc17:2019}, but this dataset is likely part of GPT 3.5's training data.} It's worth noting that while \sysname exhibits strong performance on this dataset, its proficiency may vary across diverse datasets. \textit{Finally, the current \sysname iteration does not delve into interactive visual encodings.} Our exploration revealed that blending data operations with visual analysis in different languages can be counterproductive. While we opted for static visualization libraries for consistency, future iterations of \sysname might bridge this gap, tapping into the potential of interactive visual libraries.

\subsection{Future Work}\label{future_work}
Our findings point to several fruitful areas of future research for visual analytic chatbots:

\vspace{1mm}
\noindent{\textbf{Enhancing Conversational Context Management:}} While extended token limits in LLMs allow for richer conversational contexts, they might not suffice for analytic dialogues, which exhibit unique patterns distinct from typical conversations. In \sysname, we employed a user-driven transition between the \mchat and \thread for context management. However, deep into an analysis users may also forget to open a thread. The potential exists for implicit transitions, similar to the Wizard in~\cite{tory:tc17:2019}. There are important design trade-offs between explicit (click on button) and implicit (inferring from utterance) initialization of a thread and focusing on the context provided to the chatbot. Exploring these different design trade-offs is out of the scope of our present work. Balancing the merits and challenges of explicit versus implicit context transitions, and exploring ways to refine these transitions based on user behavior, present fertile grounds for exploration.

\vspace{1mm}
\noindent{\textbf{Evolving Conversational Interfaces for Analysis:}} Feedback from usability interviews (Section~\ref{user_interviews}) indicates a need to advance the conventional chat interface for data analysis. Participants sought enhanced control over the analytic dialogue flow and its evolution. Leveraging existing research on visual analytic provenance~\cite{Gotz:AnalyticProvenance:2008,Ragan:AnalyticProvenance:2008} can improve chatbot user experiences. Further, integrating features like analysis suggestions~\cite{Srinivasan:Snowy:2021} can further enhance the interactive dimension and collaborative potential of chatbots like \sysname. Finally, participants in the usability interviews emphasized the importance of sharing their analyses with others. While~\sysname allows analyses to be stored and revisited, being able to effectively extract and organize parts of the conversation to share remains an open challenge. Situating AI chatbots with the larger sociotechnical considerations of their use is essential for guiding further explorations in this area~\cite{Gkingo:AI_workplace:2023}.

\vspace{1mm}
\noindent{\textbf{Instituting Analytic Chatbot Guardrails:}} The persuasive capabilities of computer technology~\cite{Fogg:PersuasiveComputers:1998} and visualization systems in particular~\cite{Pandel:Vis_Persuasian:2014} have long been known. However, they should be re-examined with greater urgency in light of AI analytic agents. Hallucinations with LLMs continue to be a challenge and these issues extend to data analysis and visualization. ~\sysname appeared to have reasonably high accuracy (e.g., analyzing the correct columns, executing appropriate data transformations, generating the appropriate encoding, and carrying out correct modifications). However, it also failed in ways that were not immediately obvious, such as changing the underlying data in the chart (Figure~\ref{fig:participant_teaser}, P40) or providing an incorrect explanation (Figure~\ref{fig:prolific_explain}, P35)). Instances of these errors are interleaved with reasonable and correct responses from~\sysname, and as such, could be easily overlooked. They also did not occur consistently, as some sessions had few or no errors while others had more. Developing guardrails that monitor these issues and address them is important for ensuring the accuracy of analyses and the trustworthiness of analytic chatbots.

\section{Conclusion}\label{conclusion}
The program synthesis capabilities of Large Language Models provide new opportunities and challenges for the construction of analytic conversational interfaces. We designed~\sysname to explore some of these preliminary challenges concerning the management of context and the preservation of conversational coherence. Through user studies and a usage scenario, we show that~\sysname is capable of supporting both the initial creation and refinement of visual encodings. Moreover, we show that its capabilities can support a richer conversational dialogue that alternates between text and visuals. While~\sysname and LLMs, in general, are not without their limitations, our findings point to their compelling potential and important avenues for future work.

\bibliographystyle{ACM-Reference-Format}
\bibliography{main}

\newpage
\appendix
\section{Analytic Conversations with Chatbots: Additional Details}\label{appendix_formative_study}

We re-analyzed the results for 21 participants pertaining to the \textit{context-only} and \textit{context + intent} conditions from  by Tory and Setlur (2019)~\cite{tory:tc17:2019} to establish a baseline of visual analytic conversations with AI chatbots. Tory and Setlur summarize their results as a state transition model for analytic conversations and the importance of these transitions to maintain visual and conversational coherence. Compared to their analysis, we focus more on understanding the overall structure and evolution of these conversations to assess the responses AI agent that is based on a GPT language model.
 
\subsection{The Structure of Analytic Conversations}\label{app:analytic_convo_structure}

\subsubsection{Visualizing Discourse Chunks.} \label{app:disourse_chunks}

We translated these linear conversations into a branching tree structure based upon utterance markers, such as `start over` or `reset', that separate discourse chunks. In Figure~\ref{fig:tc17_chain_comparison} we show the structure conversational structure of four participants; additional conversation trees are in the online supplemental materials. Each branch is one discourse chunk and the start of each branch is a new analytic question. These conversations contain analytic questions with branching structures that representing refinements of the data attributes (e.g., adding or removing attributes), data operations (e.g., filtering), or visualization (e.g., changing the encoding type; binding attributes to different channels).

\subsubsection{Classifying Utterances.}\label{app:utterance_classification} 

\begin{figure}[ht]
\begin{center}
    \includegraphics[width=0.48\textwidth]{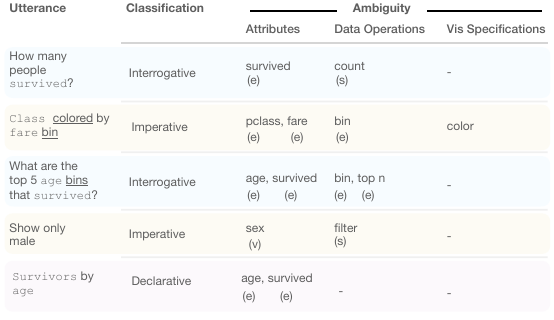}
  \end{center}
  \caption{Examples utterances according to their phrasal type and ambiguity. We use the definitions from~\cite{Srinivasan:NLI:2017} to define reference ambiguity to \texttt{attributes}, \underline{data operations}, or \underline{encoding choices}. We also extend the definition of ambiguity to include reference to a prior visualization} 
  \label{fig:utterance_classification}
\end{figure}

Prior research  by Gao~\textit{et al.}~\cite{Gao:DataTone:2015} and Srinivasan~\textit{et a.}~\cite{Srinivasan:NLI:2017} have proposed classifications for conversational utterances and their ambiguity. We find the utterances from this study resemble findings align with the classifications (questions, commands, queries, and others) proposed in~\cite{Srinivasan:NLI:2017}, however, we recast these definitions here to align with linguistic classifications of utterances in discourse~\cite{Craige:discourse:2018,kuijpers:mmi2:1992}. We propose using \textit{interrogative} in lieu of questions and \textit{imperative} in lieu of commands. While less uncommon, some  utterances took the form of declarative utterances (queries or other), such as \textit{``Women in first class who survived by age''} or \textit{``survivors by age''}, that are posed neither as a question nor a specific instruction to do something. Even though the imperative ``show me'' command could be implied (e.g., \textit{``show me survivors by age''}), we classify the utterance as declarative because ``show me'' is not explicitly stated.  Similar to prior research, we define ambiguity according to the utterance's specification of the attributes, data operations (including aggregations and filtering), and visualization specification (including chart types and encoding choices). Again, we use the classifications of Srinivasan~\textit{et al.}~\cite{Srinivasan:Utterance:2021} to indicate whether attribute references are explicitly matched all or portion of an attribute's name (e.g.,\textit{plot class by age}), are semantically similar (e.g.,~\textit{``on average how much did people pay?''}; paid is semantically similar to fare), based upon values (e.g.,~\textit{``Show only male'', male it a value in the sex attribute}), or is implicit based upon the response (e.g.,~\textit{``bin by 10-year intervals''} when age is the chart's x-axis).

\subsubsection{The evolution of analytic conversations.}\label{app:analytic_convo_evo} 

\begin{figure}[ht]
    \centering
    \includegraphics[width=1\textwidth]{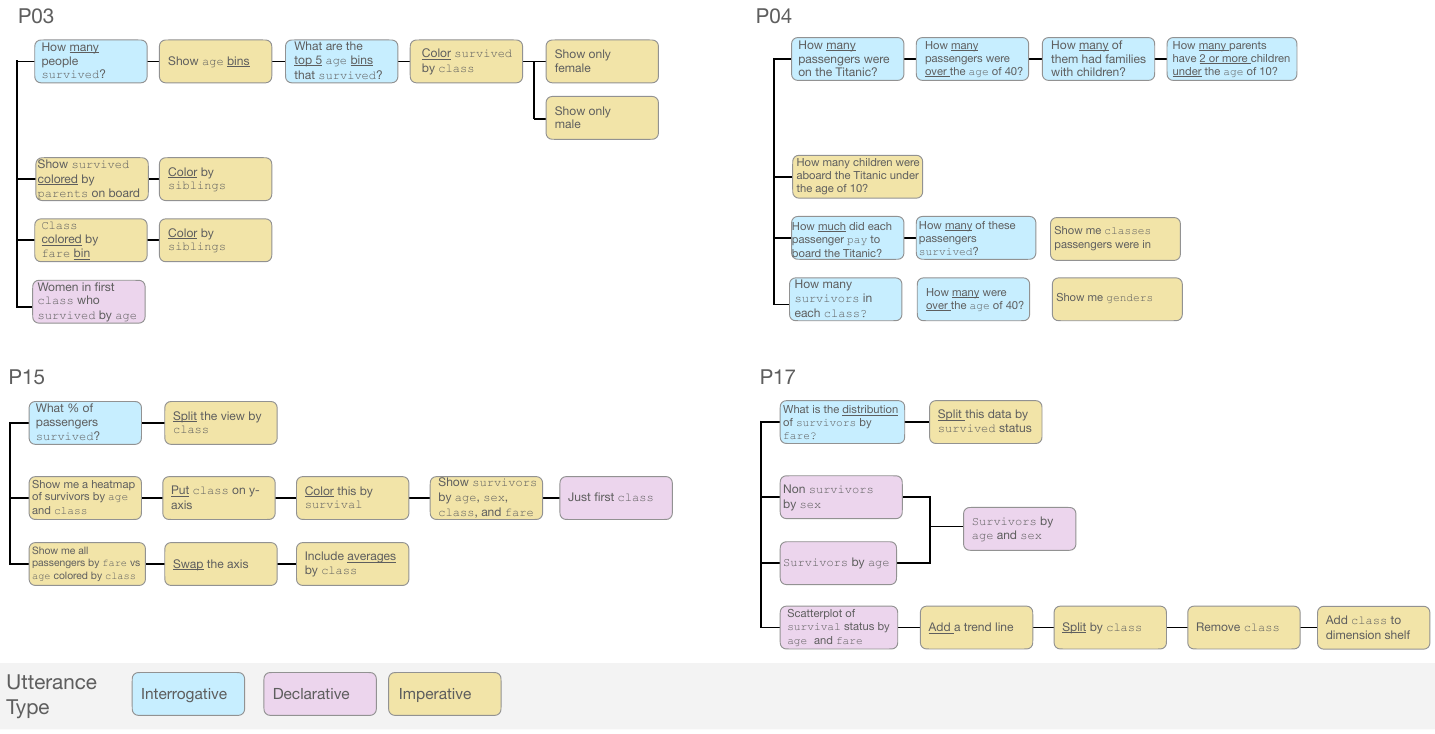}
    \caption{Visualizations of discord chunks derived from participants in~\cite{tory:tc17:2019}.}
    \label{fig:tc17_chain_comparison}
\end{figure}

We combined the branching structures together with the classification of individual utterances to examine the evolution of analytic conversations from this dataset. The individual approaches of participants varied, with some tending to pose more interrogative utterances throughout their sessions. However, participants typically began with interrogative utterances and then proceed imperative utterances within a branch (discourse chunk). Participants also tended to use interrogative utterances at the beginning of sessions but eventually transition to more imperative, and sometimes declarative, utterances to pose new questions (start a new branch).  Generally, all questions made reference to some data attributes (often explicitly or in a semantically similar way) and some data operation. It was less common was the specify a specify chart type and more common to refine existing charts wizard-generated by binding additional attributes to available encoding channels. Within a branch, the majority of participants used unspecified imperative utterances. These utterances only make sense when taking into account the conversational context of the specific branch. In only two instances did we observe the explicit merging (P17, Figure~\ref{fig:tc17_chain_comparison}) or splitting (P03) of two branches; the majority of branches went from the more general to a specific in a linear manner.

\subsection{Replication in ChatGPT Advanced Data Analysis}\label{app:chatgpt_repeat}
We compare the responses from the WoZ study to those produced by the ChatGPT Advanced Data Analysis (ADA) plugin. We identified the following key differences.

\vspace{1mm}
\noindent\textbf{Text vs. visualization default responses.} The WoZ study defaulted all responses to visualizations, whereas ADA tended to default to textual responses unless prompted to produce a visualization. However, ADA's behavior could be modified by providing instructions at the beginning of the session, for example instructing it to  \textit{``provide all responses as data visualizations by default and where appropriate''}\footnote{In August a custom instructions option was provided as well that could prompt ADA to response using visualizations by default and across all conversations.}. Alternatively, participants' utterances could make explicit references to creating a visualization in the imperative (e.g., \textit{``show me a visualization of [...]''} or simply \textit{``show me''} or interrogative forms (e.g., \textit{``can you show me a visualization...''}) When ADA provided a textual response it was also possible to ask it to provide a visualization as a follow-up utterance. This interplay between text and visualizations is absent in the WoZ study but is something ADA is capable of supporting.  While some prior research~\cite{Hearst:ChartChat:2019,Fast:Iris:2019} examined this interplay, the data from these studies is limited. 

\vspace{1mm}
\noindent\textbf{Phrasal classification of utterances does not impact ADA's ability to respond.} Generally, we found that it did not matter whether an utterance was posed to ADA was in an interrogative form or not (e.g., \textit{``how many people survived?''} vs. \textit{``total number of people that survived''}). Even constructing non-interrogative sentences that make little sense in the context of language dialogue still yielded a correct response(e.g. inputting just \textit{``many people survived''} without any context but the dataset). While out of the scope of this study, it would be interesting to explore this observation further since we see evidence that auto-regressive agents like  ADA do not require utterances resemble human-to-human dialogue. That is, humans may converse with an AI agent in a manner similar to conversing with other humans, but, this is not strictly necessary for an AI Agent to respond appropriately.

What did matter more were references to attributes, data operations, and encodings in the utterance. Even the terse \textit{``many people survived''} arguably contains references to a count operation (semantically similar to many) and a data attribute ('survived') compared to a more general and imperative utterance such as \textit{``summarize survival''} where summarize can refer to many data operations, such as counting, computing a proportion, computing a rate, etc. On this dataset, we found that, in general, ADA correctly mapped explicit, semantic, and value-based references to attributes. It also did so for the majority of data operations requested by participants (binning, counting, averages). When it was unsure of the mapping it would prompt for further clarification, but this was not consistent behavior sometimes it also produced an error. 
Despite the non-deterministic behavior of GPT model outputs, it was also possible to provide a non-ambiguous imperative response (\textit{``Scatter plot of survival status by age and fare''}). While there was still variability in ADA's overall it produced the expected output. 

\vspace{1mm}
\noindent{\textbf{Difficulties incorporating context appropriately.}} ADA was not able to consistently incorporate the contextual dependency of utterances. For example, P03 in~\cite{tory:tc17:2019}  has a chain of utterances that elaborate from the overall survival to modifying the chart by including age than gender (Figure~\ref{fig:tc17_chain_comparison}). ADA treats these utterances as independent and does not further amend the visualization to add age and gender to the survival. The result is that instead of incremental refinements, it produces a single chart for each utterance.  We can improve the performance if we add an anaphoric reference such as `this' or `the previous visualization' to the utterances, for example, the utterance \textit{``how many survived''}  followed by \textit{``add age bins to this''} produces a comparable visualization to the WoZ results. We observed that after several visualizations related to one chart,  it is not always necessary to include a reference term. This behavior is important as the majority of analysis refinements are ambiguous imperative statements that are dependent on the context of the discourse block. 

When ADA learned to refine a chart without needing to have an anaphoric reference it became necessary to provide explicit discourse markers to begin new analysis or visual encoding. For example, if refining a chart, it was possible to provide a `start over' essentially shifting the agent's focus to something new. However, the overall behavior of ADA was unpredictable because it was not clear how an utterance would be processed until the output was displayed. In circumstances when the dependency of utterances was interpreted incorrectly, it was necessary to either begin the analysis a new or provide an independent imperative prompt (e.g. \textit{``show me a visualization of survival status according to the top five age bins in male passengers''})

Lastly, ADA can switch context to answer questions outside of the dataset. For example, there are facts about the Titanic, such as the name of its captain, the number of rescue boats, or the name of the rescue ship, that might be of interest but are not directly in the dataset. This was not tested in the WoZ study.

\vspace{1mm}
\noindent{\textbf{Similar visualizations between WoZ and ADA.}} The visualizations that ADA produced were primarily bar charts and the same was true for the WoZ. One notable difference was that the Tableau system used in the WoZ study tended to use facets, whereas ADA did not. By default ADA either created compounded variable names (e.g., women-1, indicated females that survived) or would chose to filter the data and only display the subset. Finally, ADA also provided textual descriptions and interpretations of charts that it produced, which was absent in the WoZ study.  

\noindent{\textbf{Errors interrupt the conversational flow.}} On occasion, the WoZ study reached error states resulting from either unclear on incomplete utterances or from errors with the Tableau Software used to generate the visualization. However, these were less common. In contrast, ADA frequently produced errors when generating a visualization, once again related to utterance clarity but also because the code it generated was not executable. In this latter case, ADA would automatically attempt to regenerate an appropriate response.
\section{\sysname Additional Details}\label{app_capabilities}

\subsection{Editing Actions \& Utterance Intent Examples} \label{example_edits}
\begin{figure}[th!]
    \centering
    \includegraphics{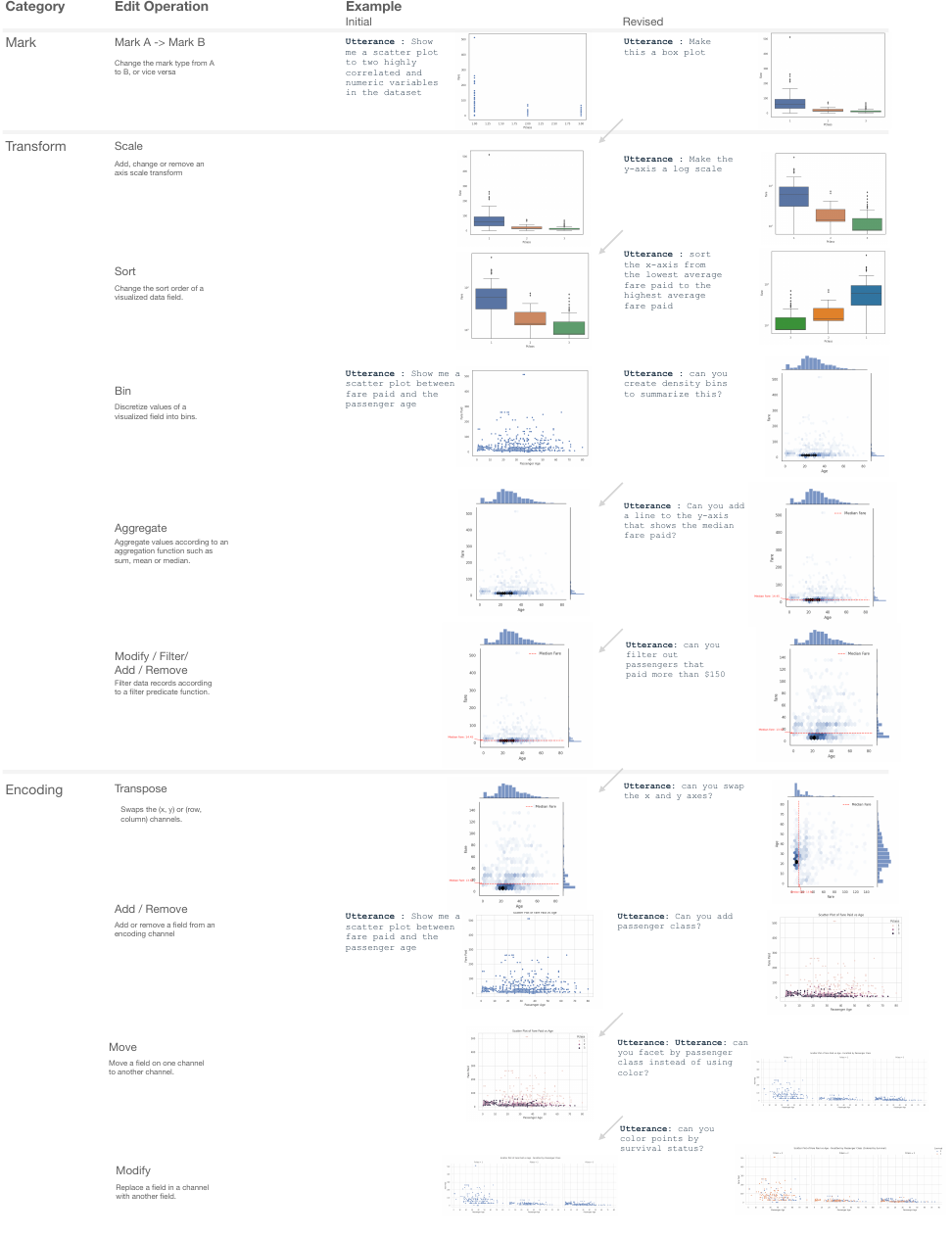}
    \caption{\textbf{Example editing operations.} The arrows indicate that the editing operation was continues in a thread. The above figure show three prompts producing a encoding that is the further modified via editing operations in a thread. Arrows indicate a continuation from the previously refined encoding. Edit operations were taken from~\cite{Kim:GraphScape:2017}.}
    \label{fig:edit-ops}
\end{figure}

\begin{figure}[th!]
    \centering
    \includegraphics[width=0.85\textwidth]{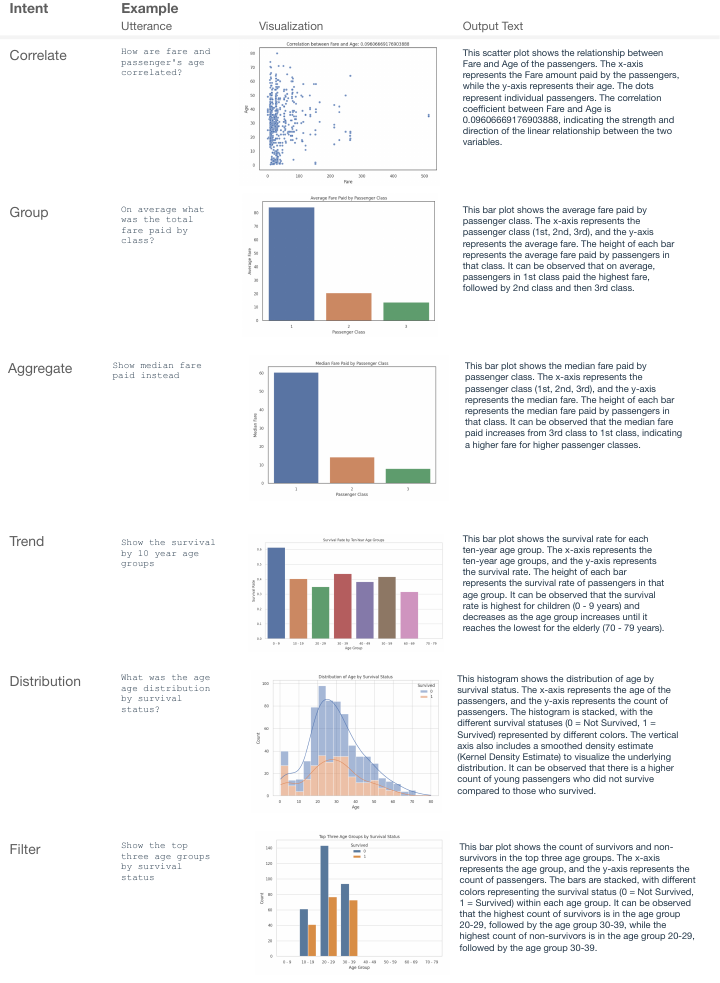}
    \caption{\textbf{Examples of~\sysname responds to utterances with different intents}. Utterances are inspired by those suggested in~\cite{Srinivasan:Snowy:2021}.}
    \label{fig:intent_example}
\end{figure}

\sysname can modify a visual encoding in a variety of ways, in Figure~\ref{fig:edit-ops} we provide a concrete set of examples from the edit operations described in GraphScape~\cite{Kim:GraphScape:2017}. Additionally, we explored how~\sysname responds utterances posed with different types of analytic intents. We draw on examples utterances from the SNOWY system~\cite{Srinivasan:Snowy:2021}. Figure~\ref{fig:intent_example} shows an example of the utterance, the resulting visualization, and a text output generated by the system to describe the data visualization that was generated. \textit{Note that due to the size of the figures, they follow the subsequent Section~\ref{app:prompt}}.

\subsection{Few-Shot Examples}\label{app:prompt}


The following are the few shot examples provided to the GPT-3.5 ChatCompletion API via~\sysname as conversational context following the system instructional prompt (Section~\ref{system:code_gen}).

During pilot studies, we noticed that GPT-3.5 tended to hallucinate semantically relevant yet non-existent colors or color palettes when generating or modifying visualizations from color-related user utterances, unless it learned how to specify colors in the demonstrated approach. Defining every variable in the examples, even the chart's \texttt{width} and \texttt{height}, before they are used also allowed the model to learn how to correctly modify charts. The model is also taught how to incorporate computations to its captions.

\vspace{1mm}

\noindent\textbf{Example 1:}

\noindent\textcolor{question}{[ROLE: USER]}
\begin{lstlisting}
How many rows are in the dataset?
\end{lstlisting}

\noindent\textcolor{response}{[ROLE: ASSISTANT]}
\begin{lstlisting}[language=Python]
```python
import pandas as pd
import seaborn as sns
import matplotlib.pyplot as plt
import base64
import io

tab10 = ["#4e79a7","#f28e2c","#e15759","#76b7b2","#59a14f","#edc949","#af7aa1","#ff9da7","#9c755f","#bab0ab"]
sns.set_palette(tab10)
sns.set(style="whitegrid")

# Load the dataset
df = pd.read_csv('./workspace/{filename}')

# Get the number of rows in the DataFrame
num_rows = df.shape[0]

# Create the bar chart using seaborn, adjusting the size of the chart as needed
width = 8
height = 2
plt.figure(figsize=(width, height))
sns.barplot(x=[num_rows])

# Add labels, annotations, and/or title
plt.xlabel('Number of Rows')

# Display the chart as a Bytes64 PNG representation with a tight layout
plt.tight_layout()
buffer = io.BytesIO()
plt.savefig(buffer, format='png')
buffer.seek(0)
png_data = base64.b64encode(buffer.getvalue()).decode('utf-8')
buffer.close()

plt.clf()
plt.close()

# Print the chart image
print(png_data)

# Print a caption
print("This plot shows a single bar representing the total number of rows in the dataset, which is " + str(num_rows) + ".")
```
\end{lstlisting}

\vspace{1mm}

\noindent\textbf{Example 2:}

\noindent\textcolor{question}{[ROLE: USER]}
\begin{lstlisting}
Plot the distributions of every continuous measure.
\end{lstlisting}

\noindent\textcolor{response}{[ROLE: ASSISTANT]}
\begin{lstlisting}[language=Python]

```python
import pandas as pd
import seaborn as sns
import matplotlib.pyplot as plt
import base64
import io

tab10 = ["#4e79a7","#f28e2c","#e15759","#76b7b2","#59a14f","#edc949","#af7aa1","#ff9da7","#9c755f","#bab0ab"]
sns.set_palette(tab10)
sns.set(style="white")

# Load the dataset
df = pd.read_csv('./workspace/{filename}')

# Selecting the continuous variables from the DataFrame
continuous_vars = []

for col_name, dtype in df.dtypes.iteritems():
    if dtype==float:
        continuous_vars.append(col_name)

# Creating a new DataFrame with only the continuous variables
continuous_df = df[continuous_vars]

# Melt the DataFrame to create a long-form representation
melted_df = pd.melt(continuous_df)

# Plotting a faceted histogram
g = sns.FacetGrid(melted_df, col="variable", col_wrap=2, sharex=False, sharey=False)
num_bins = 20
g.map(sns.histplot, "value", bins=num_bins)
g.set_titles("{col_name}")
g.set_axis_labels("", "Count")

# Display the chart as a Bytes64 PNG representation with a tight layout
plt.tight_layout()
buffer = io.BytesIO()
plt.savefig(buffer, format='png')
buffer.seek(0)
png_data = base64.b64encode(buffer.getvalue()).decode('utf-8')
buffer.close()

plt.clf()
plt.close()

# Print the chart image
print(png_data)
# Print a caption
print("This plot shows a grid of histograms, each representing a different continuous variable. The x-axis represents the value range of each variable, and the y-axis represents the count of occurrences.")
```

\end{lstlisting}

\end{document}